\documentclass[
  twoside,
  twocolumn,
  prb,
  preprintnumbers,
  floatfix,
  showpacs,
]{revtex4}

\usepackage{amsmath}
\usepackage{amssymb}
\usepackage{color}
\usepackage{graphicx}
\usepackage{tabularx}
\usepackage{subfigure}
\usepackage{dcolumn}
\usepackage{times}

\newcommand{\etal}[0]{\textit{et al.}}
\newcommand{\Al}[0]{\text{Al}}
\newcommand{\Sb}[0]{\text{Sb}}
\newcommand{\AlSb}[0]{\text{AlSb}}
\renewcommand{\H}[0]{\text{H}}
\newcommand{\C}[0]{\text{C}}

\newcommand{\Ge}[0]{\text{Ge}}
\newcommand{\Sn}[0]{\text{Sn}}
\renewcommand{\P}[0]{\text{P}}
\renewcommand{\O}[0]{\text{O}}

\newcommand{\Te}[0]{\text{Te}}
\newcommand{\bulk}[0]{\text{bulk}}

\newcommand{\VBM}[0]{\text{VBM}}
\newcommand{\CBM}[0]{\text{CBM}}
\newcommand{\ext}[0]{\text{ext}}
\newcommand{\intt}[0]{\text{int}}

\newcommand{\tet}[0]{\text{tet}}
\newcommand{\hex}[0]{\text{hex}}
\newcommand{\K}[0]{\text{K}}
\newcommand{\eV}[0]{\text{eV}}
\newcommand{\meV}[0]{\text{meV}}
\newcommand{\cm}[0]{\text{cm}}

\newcommand{\AAA}[0]{\text{\AA}}

\newcommand{\sect}[1]{Sec.~\ref{#1}}
\newcommand{\fig}[1]{Fig.~\ref{#1}}
\newcommand{\eq}[1]{Eq.~(\ref{#1})}

\renewcommand{\vec}[1]{\ensuremath\boldsymbol{#1}}
\renewcommand{\epsilon}[0]{\varepsilon}

\newlength{\myhgt}
\newlength{\myskip}

\begin{document}

\preprint{published as Phys. Rev. B. {\bf 81}, 195216 (2010)}

\title{
  Extrinsic point defects in aluminum antimonide
}

\author{Paul Erhart} \email{erhart1@llnl.gov}
\author{Daniel {\AA}berg}
\author{Vincenzo Lordi} \email{lordi2@llnl.gov}
\affiliation{
  Physical and Life Science Directorate,
  Lawrence Livermore National Laboratory,
  Livermore, California, 94550, USA
}

\begin{abstract}
We investigate thermodynamic and electronic properties of group IV (C, Si, Ge, Sn) and group VI (O, S, Se, Te) impurities as well as P and H in aluminum antimonide (AlSb) using first-principles calculations. To this end, we compute the formation energies of a broad range of possible defect configurations including defect complexes with the most important intrinsic defects. We also obtain relative scattering cross strengths for these defects to determine their impact on charge carrier mobility. Furthermore, we employ a self-consistent charge equilibration scheme to determine the net charge carrier concentrations for different temperatures and impurity concentrations. Thereby, we are able to study the effect of impurities incorporated during growth and identify optimal processing conditions for achieving compensated material. The key findings are summarized as follows. Among the group IV elements, C, Si, and Ge substitute for Sb and act as shallow acceptors, while Sn can substitute for either Sb or Al and displays amphoteric character. Among the group VI elements, S, Se, and Te substitute for Sb and act as deep donors. In contrast, O is most likely to be incorporated as an interstitial and predominantly acts as an acceptor. As a group V element, P substitutes for Sb and is electrically inactive. C and O are the most detrimental impurities to carrier transport, while Sn, Se, and Te have a modest to low impact. Therefore, Te can be used to compensate C and O impurities, which are unintentionally incorporated during the growth process, with minimal effect on the carrier mobilities.
\end{abstract}

\maketitle

\section{Introduction}

Applications in nuclear and medical imaging as well as homeland security have recently led to a resurgence of research in the field of radiation detection. Specifically, this has motivated the search for new materials to improve the performance of room-temperature semiconductor radiation detectors.\cite{Owe06, LukAmm07, May67} Aluminum antimonide, which is a III-V semiconductor with an indirect band gap of 1.69\,eV, is a promising candidates but its potential for radiation detection has not been conclusively established yet.\cite{ArmSwiShe77, KutRybAby01}

The basic mechanism underlying radiation detection in a semiconductor can be summarized as follows:
(1) an incoming quantum of radiation transfers energy to the electrons in the material creating electron-hole pairs,
(2) the charge carriers are separated in an externally applied electric field,
(3) the magnitude of the resulting electric current is measured.
To achieve maximum energy resolution, the ``loss'' of charge carriers due to scattering and recombination events must be minimized. This requires high carrier mobilities and very low concentrations of intrinsic defects and impurities. In addition, a very high resistivity is desirable to maintain a low concentration of background free carriers and thereby to minimize noise events. For instance, in germanium detectors that are used at cryogenic temperatures, defect and impurity concentrations have to be reduced to less than $10^{10}\,\cm^{-3}$ and free carrier densities to less than $10^{14}\,\cm^{-3}$ to reach optimal performance.\cite{Kno00} To meet such stringent bounds a highly optimized manufacturing process is needed, which typically implies years of research and development. This situation renders the screening of candidate materials extremely challenging. Predictive computations, however, offer a possibility to circumvent this difficulty, since they allow studying the fundamental limits of materials in a much more time effective and systematic fashion. Specifically, quantum mechanical methods based on density-functional theory can be used to compute the formation energies of intrinsic and extrinsic defects and their equilibrium concentrations. They also allow one to obtain the defect-limited scattering rates, which determine the mobility and life time of free charge carriers. Since these calculations can be done for idealized systems, it is possible to separate the effects of different intrinsic and extrinsic defects and determine the ultimate performance limits.

The objective of the present paper is to elucidate the role of extrinsic defects in aluminum antimonide on the basis of first-principles calculations. Specifically, we seek to establish which elements are the most detrimental for the electronic properties of AlSb and which elements can be introduced intentionally to compensate for intrinsic defects and impurities. In short, it will be shown that
({\em i}) C, Si, and Ge substitute on Sb sites and act as acceptors with shallow transition levels,
({\em ii}) Sn can substitute on both Al and Sb sites and displays amphoteric characteristics,
({\em iii}) S, Se, and Te substitute on Sb and act as donors with deep equilibrium transition levels,
({\em iv}) O is most likely to be incorporated as an interstitial and predominantly acts as an acceptor,
({\em v}) P substitutes for Sb and is electrically inactive,
({\em vi}) C and O have the most detrimental effect on carrier transport, while
({\em vii}) Te is a weak scatterer and can therefore be used to manipulate the net concentration of charge carriers with a modest effect on their mobilities. The present study represents a continuation of our earlier work on intrinsic defects in AlSb \cite{AbeErhWil08} and is connected to our recent fully quantum-mechanical calculations of defect-limited carrier mobilities.\cite{LorErhAbe10}

\begin{figure*}
  \setlength{\myhgt}{0.9in}
  \setlength{\myskip}{0.in}
  \centering
  \begin{tabular}[c]{lcl}
    \begin{tabular}[b]{rr}
      \subfigure[$X_{\Sb}$ ($C_{3v}$)]{
	\includegraphics[height=\myhgt]{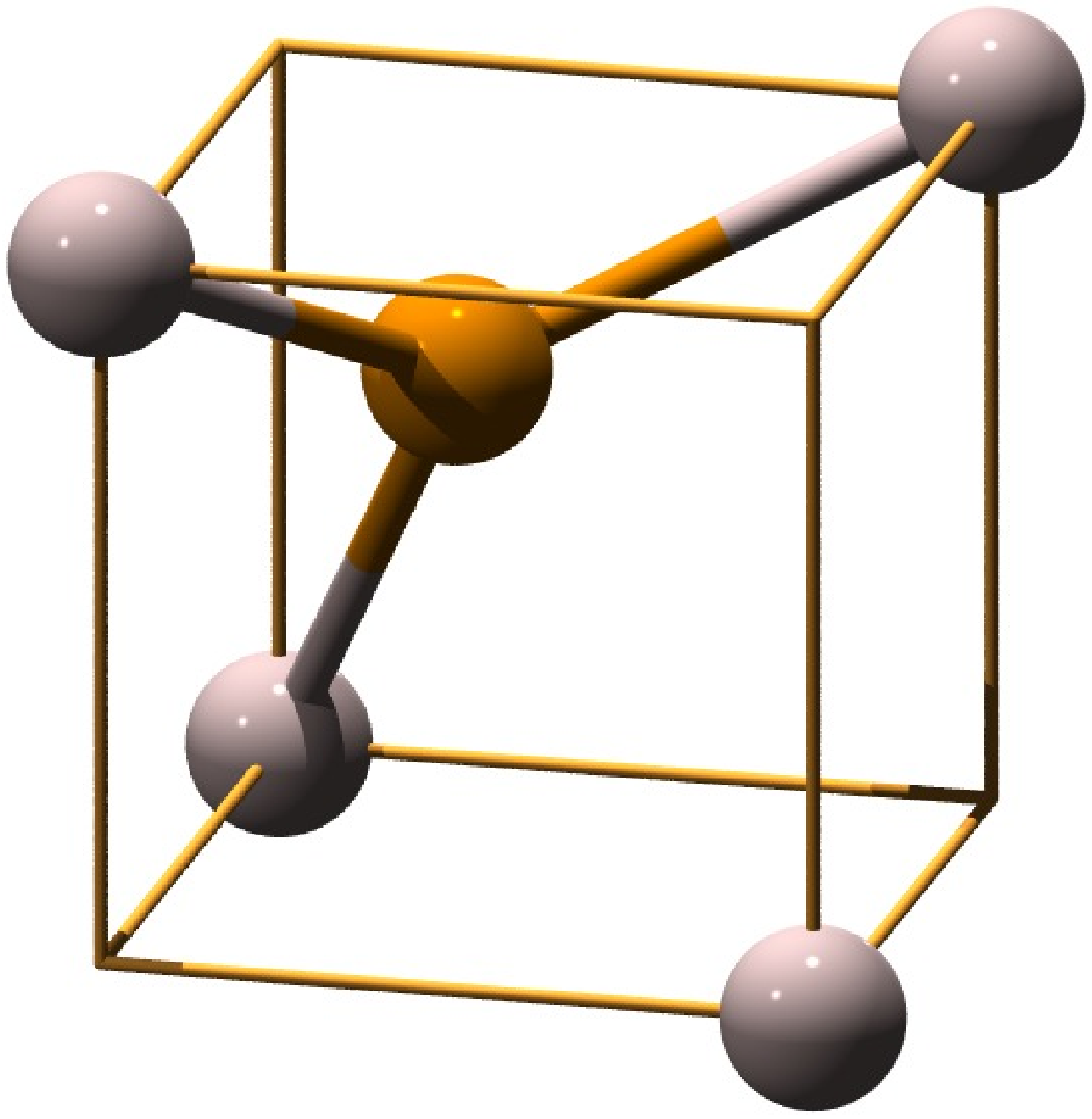}
        \label{fig:confs_X_Sb_C3}
      }
      &
      \subfigure[$X_{i,\hex}$]{
	\includegraphics[height=\myhgt]{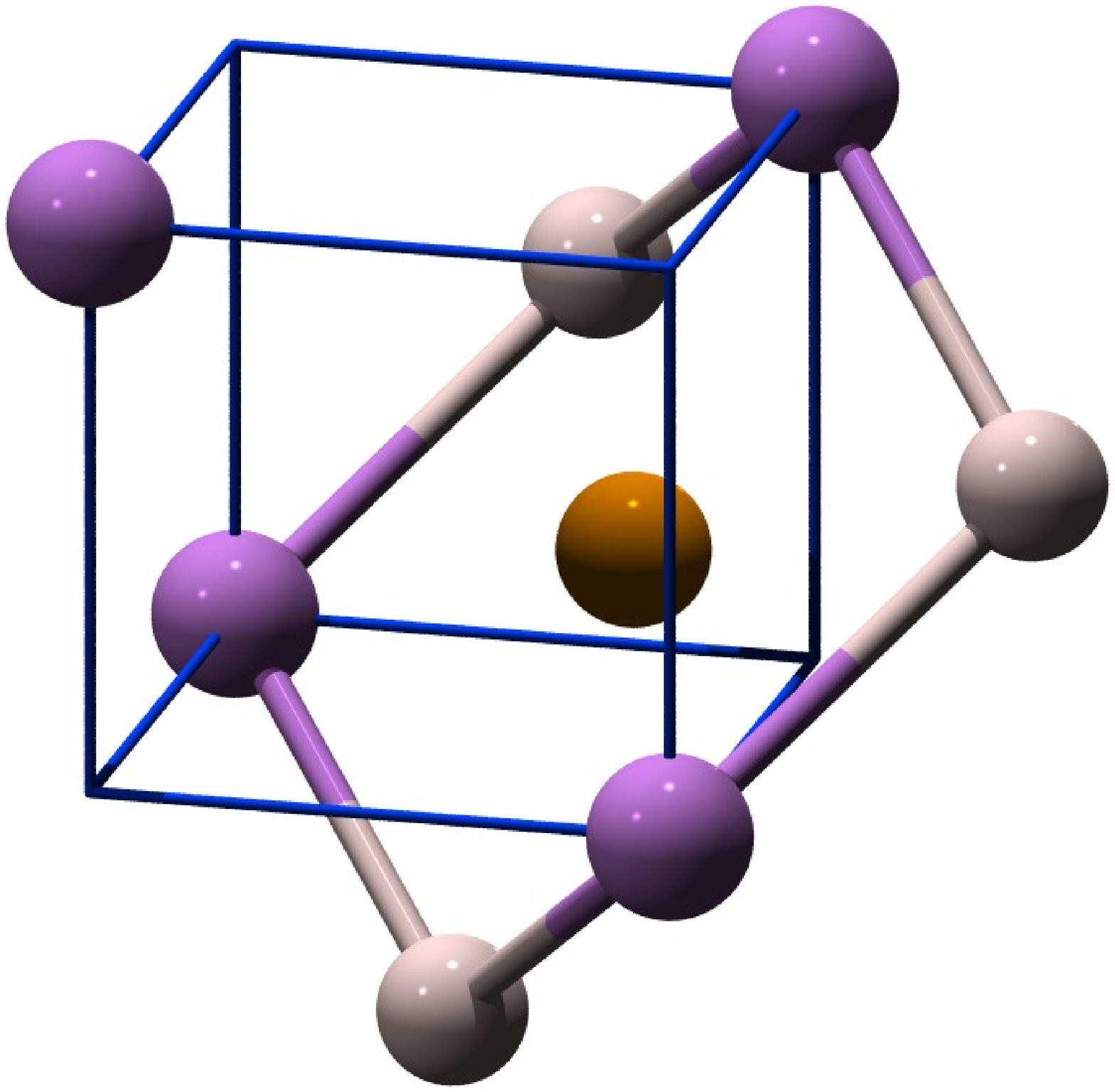}
        \label{fig:confs_int_hex}
      }
      \\
      \subfigure[$X_{i,\tet,\Al}$]{
	\includegraphics[height=\myhgt]{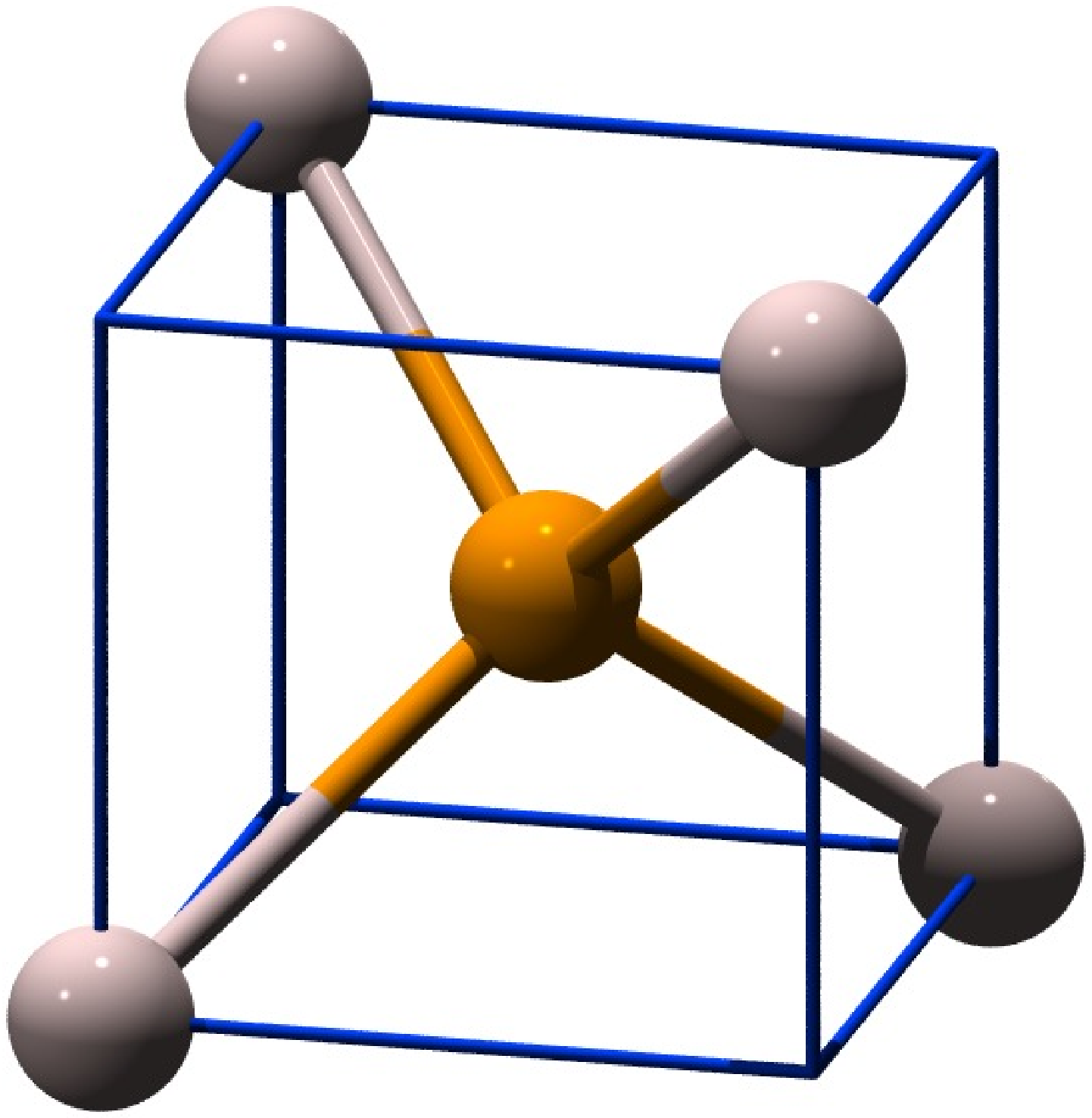}
        \label{fig:confs_int_tet_Al}
      }
      &
      \subfigure[$X_{i,\tet,\Sb}$]{
	\includegraphics[height=\myhgt]{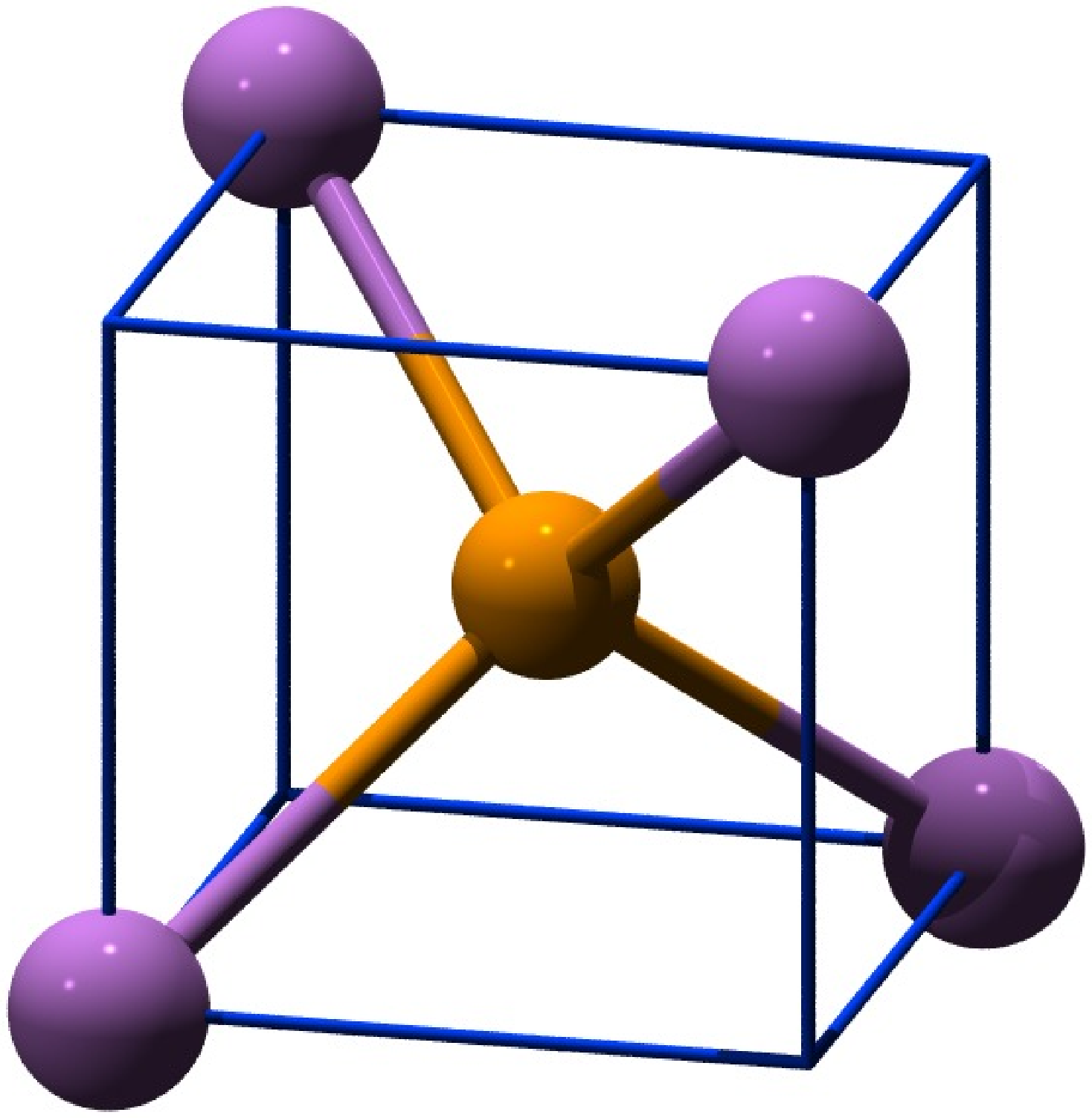}
        \label{fig:confs_int_tet_Sb}
      }
    \end{tabular}
    &
    \subfigure[ideal]{
      \includegraphics[height=2.2\myhgt]{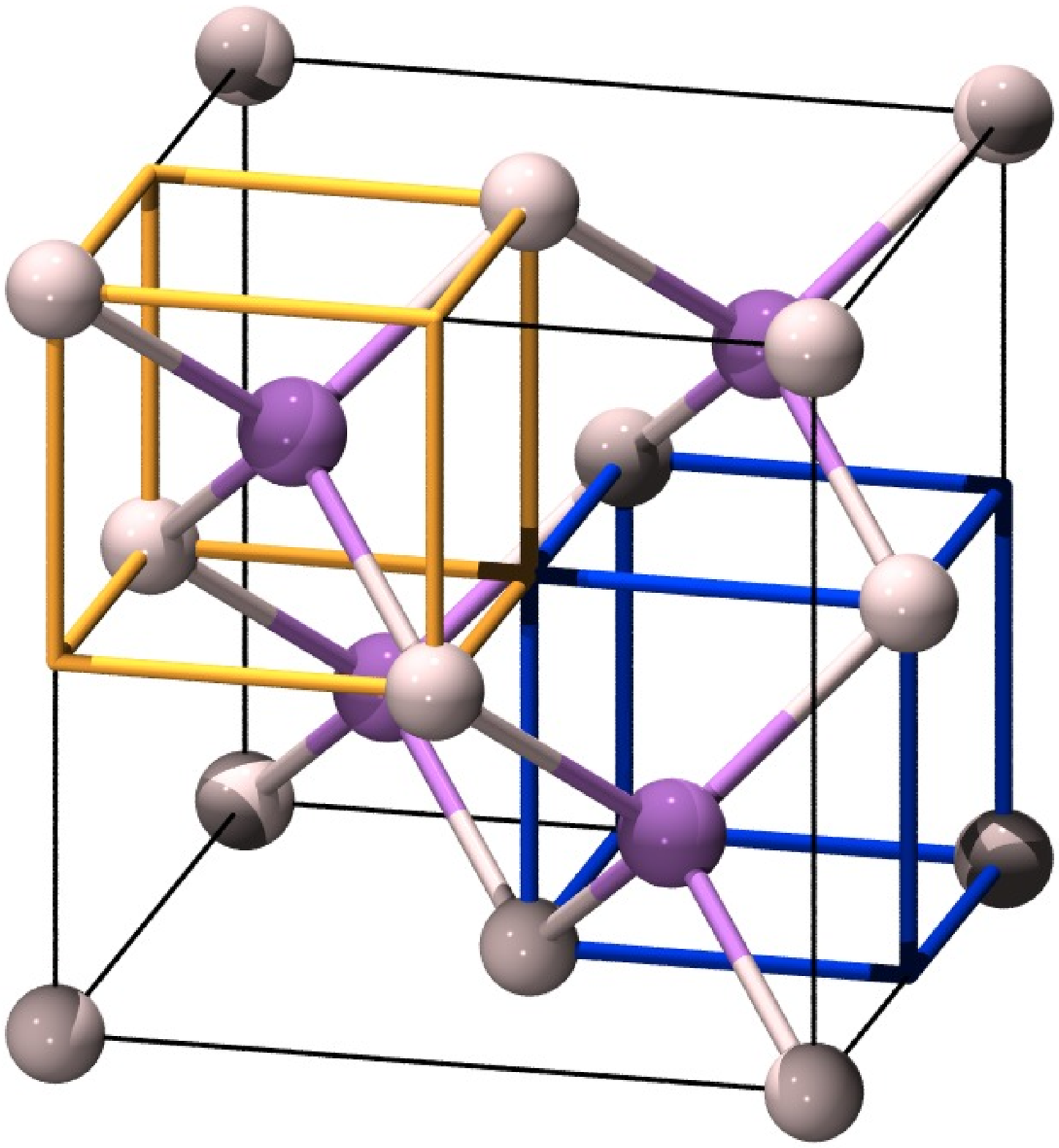}
    }
    &
    \begin{tabular}[b]{rr}
      \subfigure[$(X-\Al)_{\Al}^{\left<100\right>}$]{
	\includegraphics[height=\myhgt]{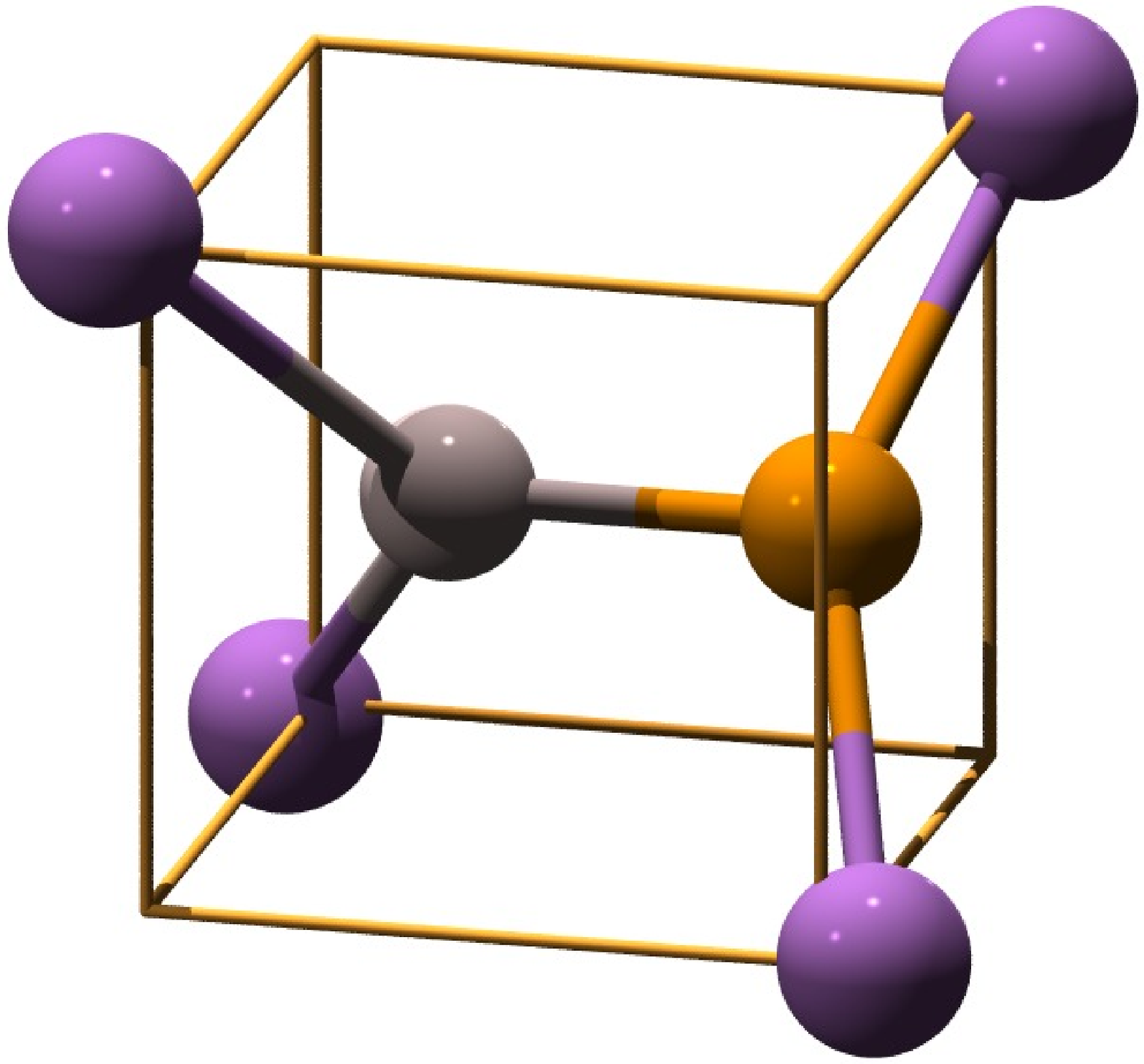}
        \label{fig:confs_XAl_Al_100}
      }
      &
      \subfigure[$(X-\Sb)_{\Sb}^{\left<100\right>}$]{
	\includegraphics[height=\myhgt]{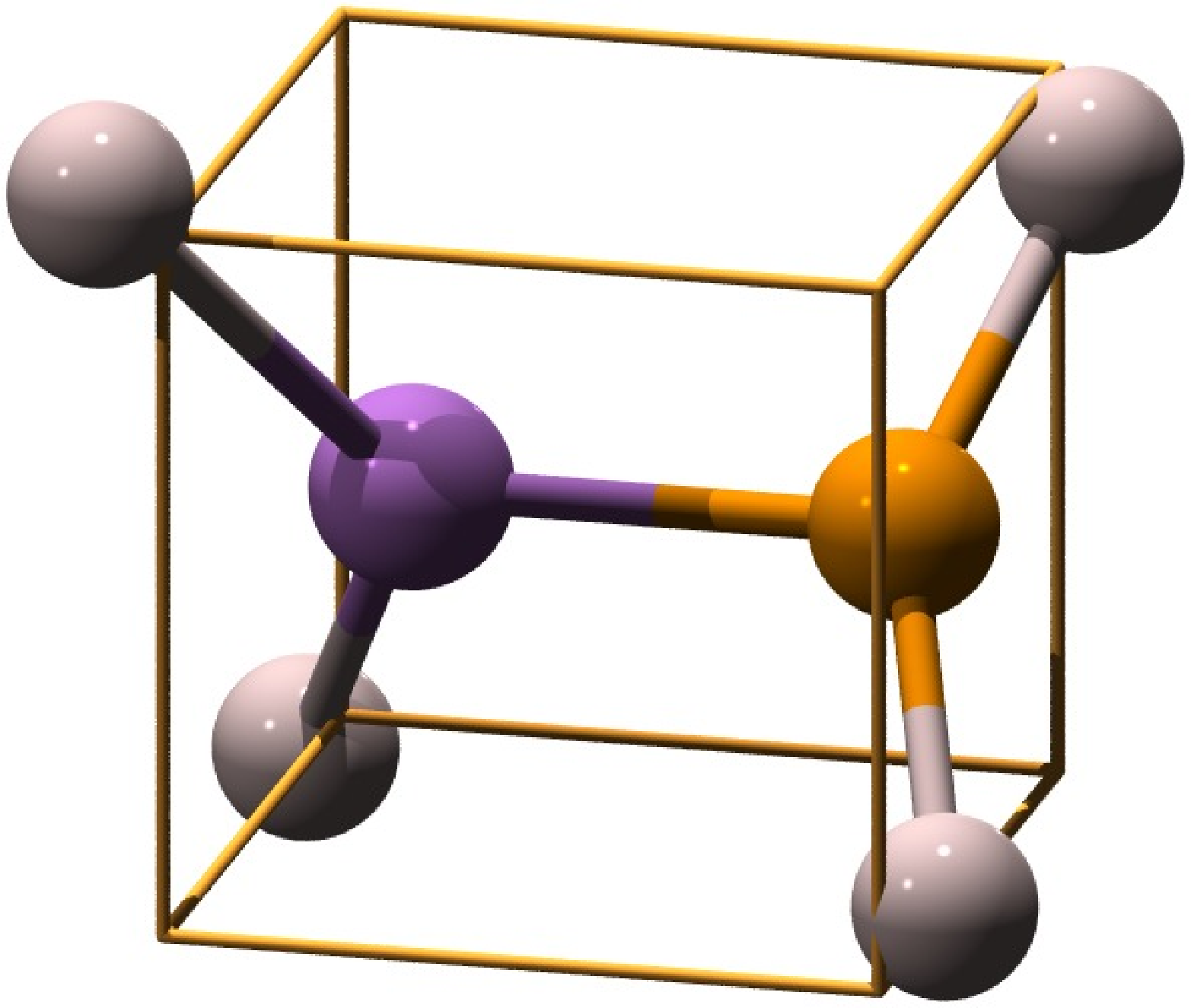}
        \label{fig:confs_XSb_Sb_100}
      }
      \\
      \subfigure[$X_{i,\text{bb}}$]{
	\includegraphics[height=\myhgt]{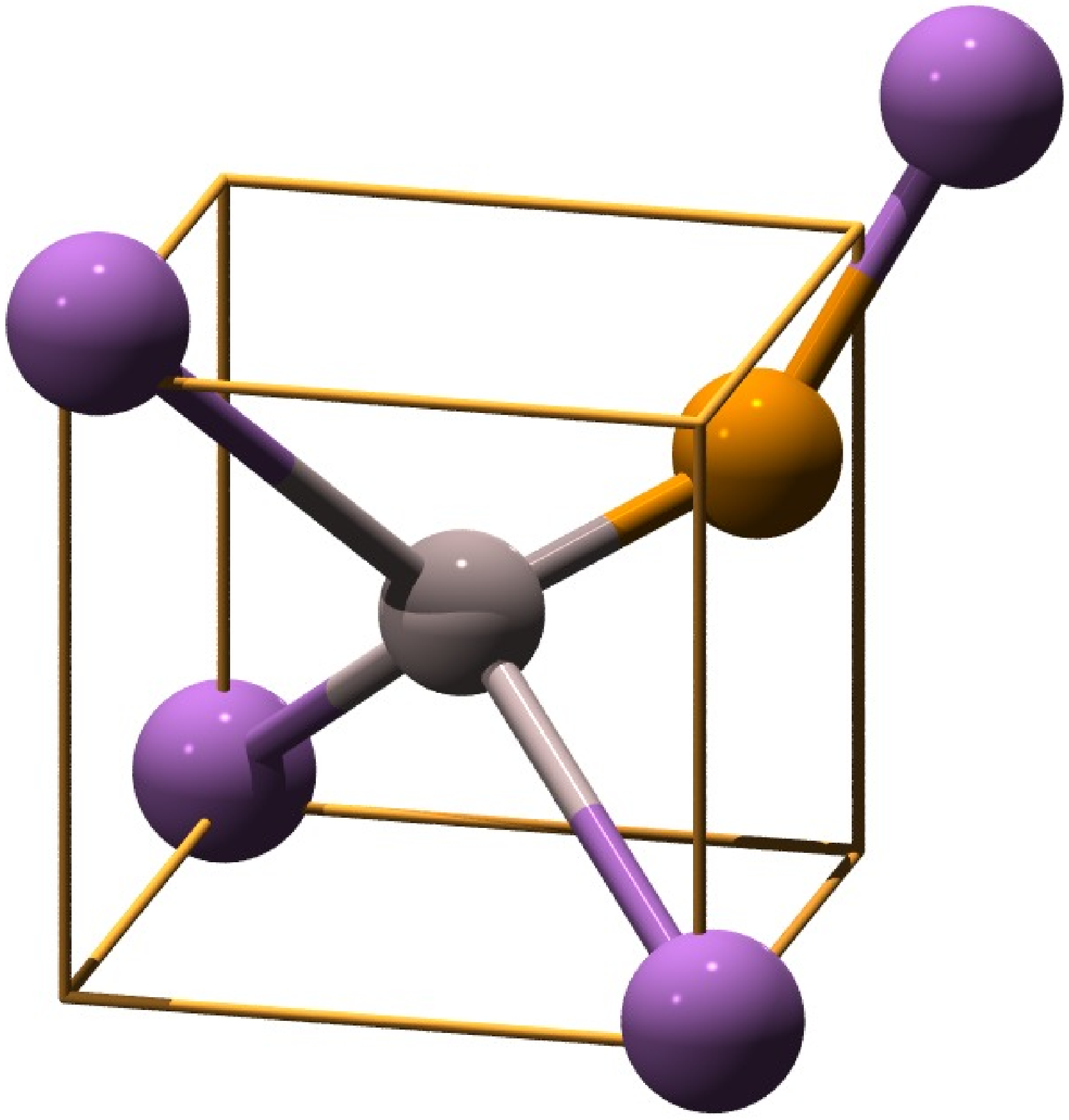}
        \label{fig:confs_int_bb}
      }
      &
      \subfigure[$\H_{i,\text{bb}}$]{
	\includegraphics[height=\myhgt]{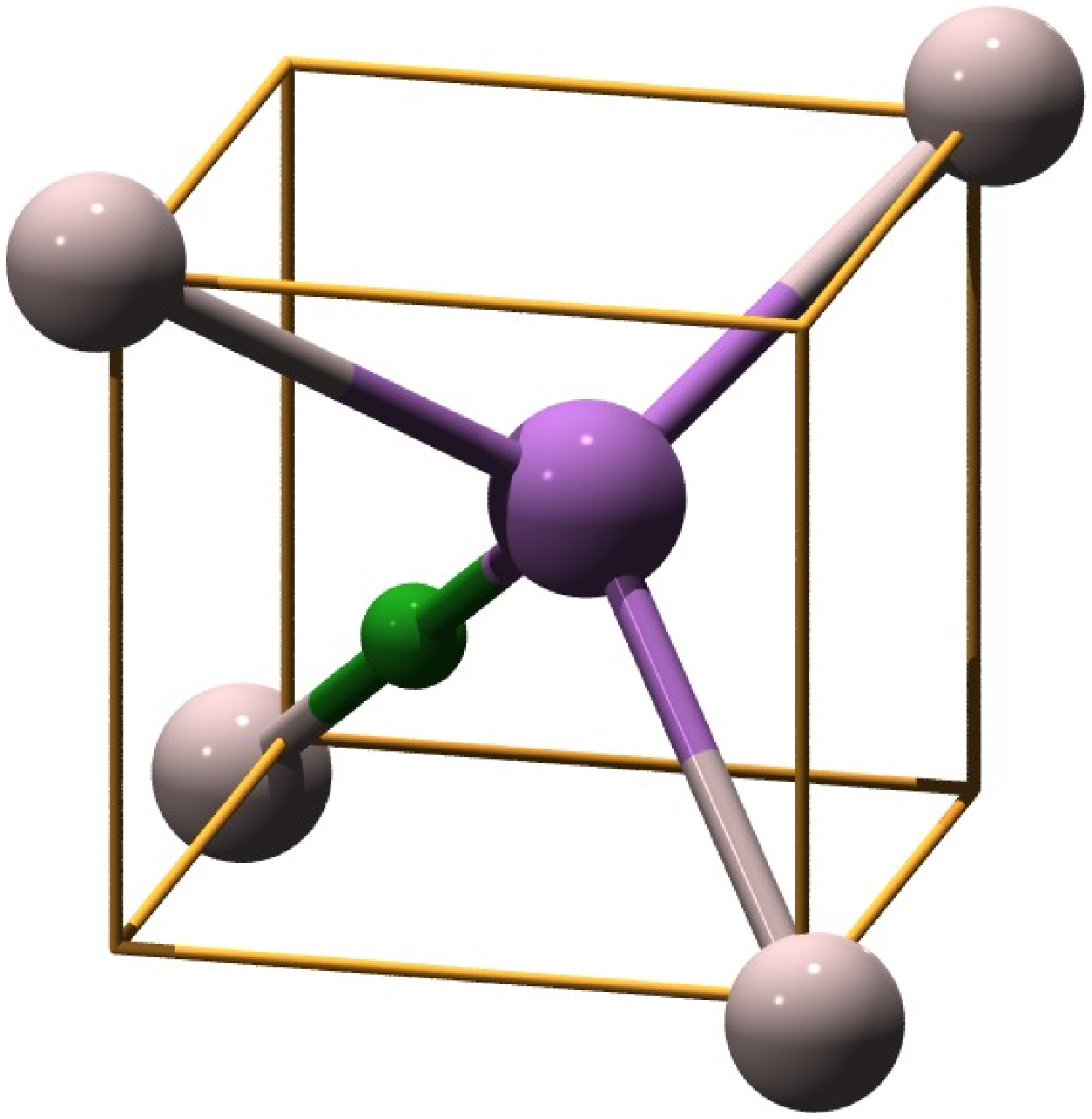}
        \label{fig:confs_int_bb_H}
      }
    \end{tabular}
  \end{tabular}
  \\
  \setlength{\myhgt}{0.85in}
  \setlength{\myskip}{0.8in}
  \subfigure[$X_{\Al}$ ($\alpha$-CCBDX)]{
    \includegraphics[height=\myhgt]{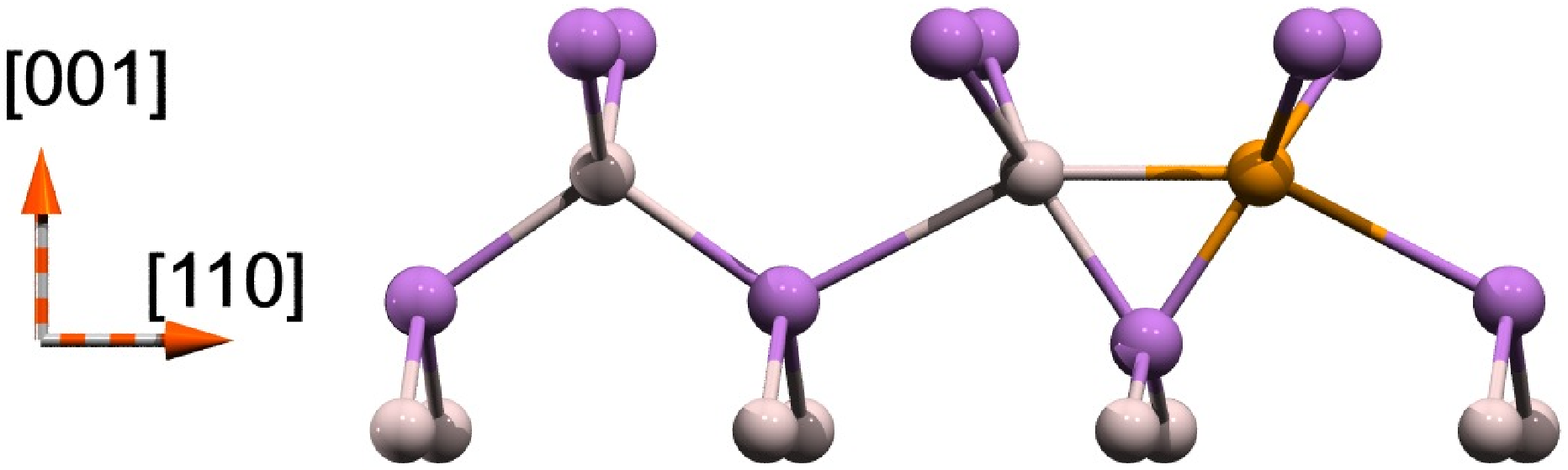}
    \label{fig:confs_X_Al_aCCBDX}
  }
  \hspace{\myskip}
  \subfigure[$X_{\Al}$ ($\beta$-CCBDX)]{
    \includegraphics[height=\myhgt]{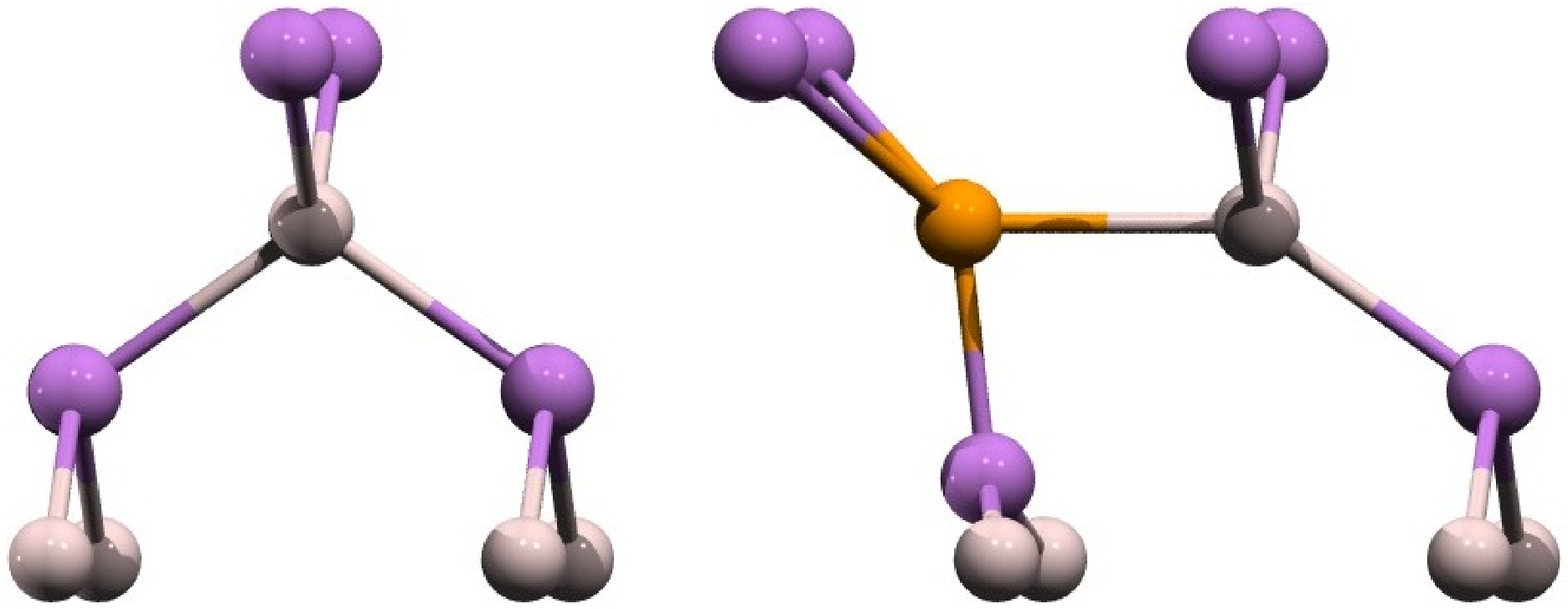}
    \label{fig:confs_X_Al_bCCBD}
    }
  \\
  \subfigure[$X_{\Sb}$ ($\alpha$-CCBDX)]{
    \includegraphics[height=\myhgt]{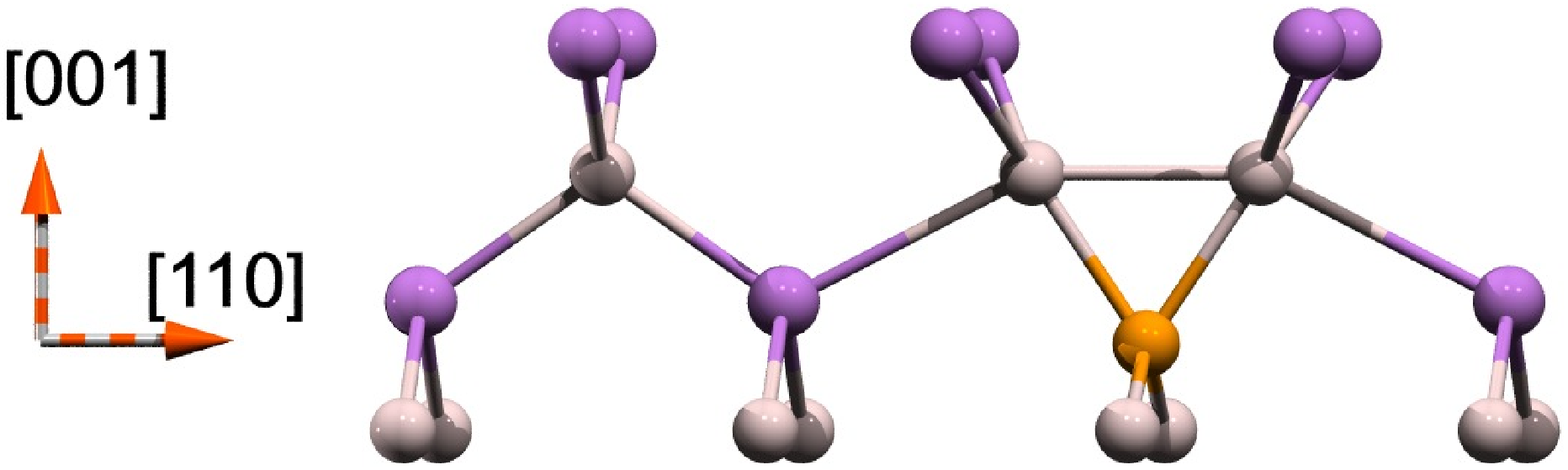}
    \label{fig:confs_X_Sb_aCCBDX}
  }
  \hspace{\myskip}
  \subfigure[$X_{\Sb}$ ($\beta$-CCBDX)]{
    \includegraphics[height=\myhgt]{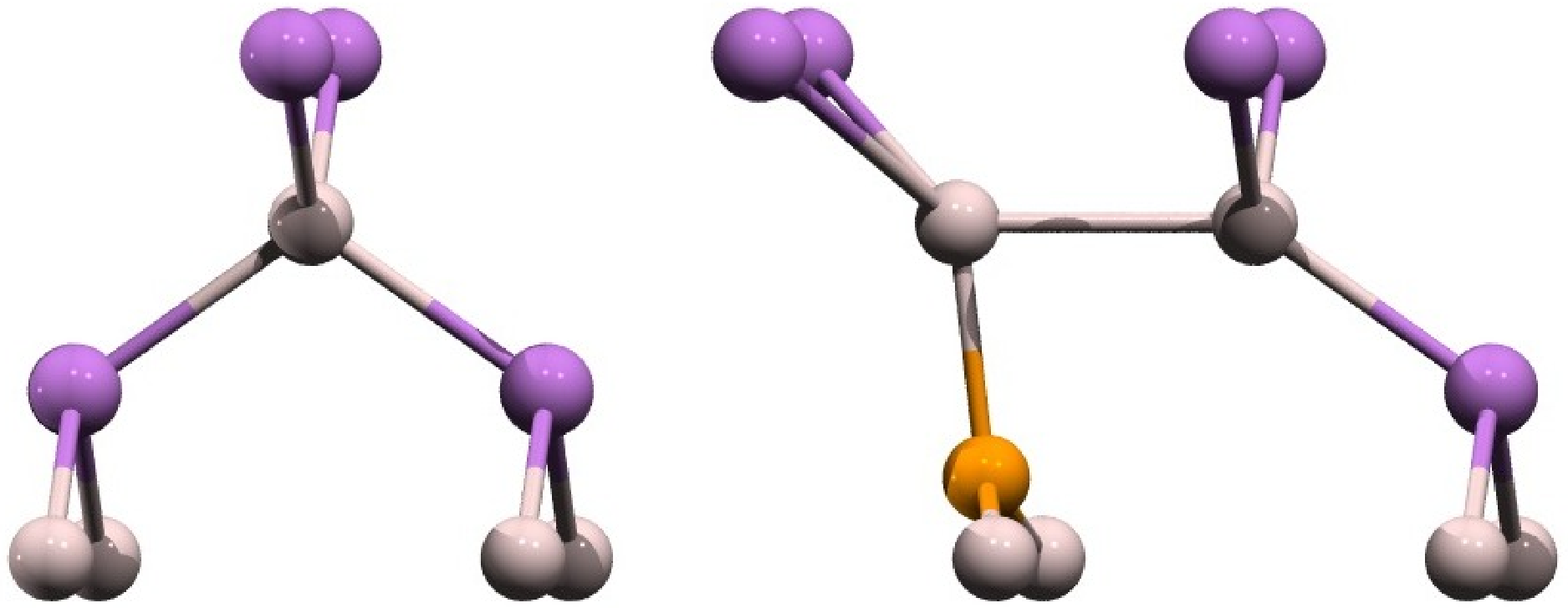}
    \label{fig:confs_X_Sb_bCCBDX}
  }
  \caption{
    (Color online) Overview of defect configurations considered in this study. Large violet, small gray, small orange and very small green spheres represent Sb, Al, impurity atoms, and H, respectively.
  }
  \label{fig:confs}
\end{figure*}

This paper is organized as follows. In \sect{sect:method} we give an overview of the methods and computational parameters. The formation energies of extrinsic defects and their binding energies in complexes with intrinsic defects are summarized in \sect{sect:eform}. Then we discuss the effect of extrinsic defects on charge carrier mobility using a measure that is based on their associated lattice distortions in \sect{sect:relscat}. Finally, we analyze the impact of selected extrinsic defects on the net charge carrier and defect concentrations in the material and discuss optimal doping schemes in \sect{sect:dneh}.

\section{Methodology}
\label{sect:method}

\subsection{Defect thermodynamics}


The equilibrium concentration, $c$, of a point defect depends on its free energy of formation, $\Delta G_f$, via
\begin{align}
   c = c_0 \exp\left(-\Delta G_f / k_B T \right),
   \label{eq:def_conc}
\end{align}
where $c_0$ is the concentration of possible defect sites. The formation free energy $\Delta G_f$ is usually approximated by the formation energy $\Delta E_f$ which is legitimate if the vibrational entropy and the pressure-volume term are small.\cite{AbeErhWil08} In the present context we are interested in impurity-related defects in a binary stoichiometric compound. In this case the formation energy of a defect in charge state $q$ can be written as \cite{ZhaNor91}
\begin{align}
  \Delta E_f
  &= E_\text{def}
  - \frac{1}{2} \left( n_{\Al} + n_{\Sb} \right) \mu_{\AlSb}^{\bulk}
  \nonumber
  \\
  & \quad
  - \frac{1}{2} \left( n_{\Al} - n_{\Sb} \right)
  \left( \mu^{\bulk}_{\Al} - \mu^{\bulk}_{\Sb} \right)
  \nonumber
  \\
  & \quad
  - \frac{1}{2} \left( n_{\Al} - n_{\Sb} \right) \Delta\mu_{\Al\Sb}
  + q \left( E_{\VBM} + \mu_e \right)
  \nonumber
  \\
  & \quad
  - \sum_i^\text{imp} \Delta n_i (\mu_i^{\bulk} + \Delta\mu_i),
  \label{eq:eform}
\end{align}
where $E_\text{def}$ is the total energy of the system containing the defect, $n_i$ denotes the number of atoms of type $i$, and $\mu_i^{\bulk}$ is the chemical potential of component $i$ in its reference state. Neglecting entropic contributions the chemical potentials of the reference phases can be replaced by the cohesive energies at zero Kelvin. The formation energy depends on the chemical environment via $\Delta\mu_{\Al\Sb}$ which describes the variation of the chemical potentials under different conditions. The range is constrained by the formation energy of AlSb, $|\Delta\mu_{\Al\Sb}|\leq \Delta H_f(\Al\Sb)$, where for the present convention $\Delta\mu_{\Al\Sb}=-\Delta H_f(\Al\Sb)$ and $\Delta\mu_{\Al\Sb}=\Delta H_f(\Al\Sb)$ correspond to Al and Sb-rich conditions, respectively. The formation energy also depends on the electron chemical potential, $\mu_e$, which is measured with respect to the valence band maximum, $E_{\VBM}$. The last sum in \eq{eq:eform} runs over all the impurities in the system, where $\Delta n_i$ denotes the difference in the number of atoms of impurity $i$ between the ideal and the defective system, and $\Delta\mu_i$ is the change in chemical potential of this component with respect to the bulk chemical potential. For the latter we used the cohesive energy of the respective most stable phase (e.g., O$_2$ molecule, Si in the diamond structure, {\it etc.}) as computed within the present computational framework.

 The formation energies that are shown in Figs.~\ref{fig:eform1}, \ref{fig:eform_O}, \ref{fig:eform_H}, and \ref{fig:ebind} have been obtained in the impurity-rich limit, i.e., $\Delta\mu_i = 0\,\eV$. We have chosen this limit because it provides a good indication for how easy a certain impurity is incorporated during the growth process. In some cases, e.g., silicon and oxygen, this leads to negative formation energies. If one is interested in formation energies under different chemical conditions, the relevant formation energies can be easily obtained by shifting all formation upwards by the value of $\Delta\mu_i$ chosen. In this way, one can e.g., determine the formation energies that are applicable if the AlSb sample is in contact with say a block of Al$_2$O$_3$.\cite{Du09}

Given the formation energies for two different charge states, $q_1$ and $q_2$, at the valence band maximum [$\mu_e=0$ in \eq{eq:eform}], we can also obtain the corresponding equilibrium (thermal) transition level,
\begin{align}
  E^{th}_{tr}(q_1\rightarrow q_2)
  = \frac{\Delta E_f(q_2) - \Delta E_f(q_1)}{q_1 - q_2}.
  \label{eq:trans}
\end{align}
Note that equilibrium transition levels $E^\text{th}_{tr}$ are distinct from electronic transition levels $E_{tr}^\text{opt}$. While the former determine the equilibrium electron chemical potential at which a change in the charge state occurs, the latter are related to optical transitions. In the present work, we obtained electronic transition levels from band structure calculations of defect cells. Examples are shown in \fig{fig:bands}.

Binding energies for defect complexes are calculated according to
\begin{align}
  \Delta E_b &= \Delta E_f^{\text{complex}} - \sum_i \Delta E_f^{(i)},
  \label{eq:ebind}
\end{align}
where the sum runs over the individual defects from which the complex is constructed. Note that $\Delta E_b$ changes with the electron chemical potential, since $\Delta E_f$ for both the complex and the individual defects depends on $\mu_e$. According to \eq{eq:ebind} negative binding energies indicate a driving force for complex formation.

\subsection{Defect structures}
\label{sect:defect_confs}

We consider C, Si, O, S, P, and H as potential impurities incorporated during the growth and include Ge, Sn, Se, and Te as possible compensating dopants. Figure~\ref{fig:confs} provides an overview of the defect configurations included in this work. To be specific, we took into account substitutional configurations with both tetrahedral ($T_d$) and trigonal ($C_{3v}$) symmetry as well as four variations of cation-cation bonded (CCBDX) configurations that have been discussed in connection with DX-centers in III-V and II-VI semiconductors.\cite{ParCha96, DuZha05} (Additional tests were carried out for several distorted substitutional configurations, including molecular dynamics simulations in order to check for the existence of other relevant local minima, but did not reveal any further configurations of importance). With regard to interstitial configurations, we included tetrahedral ($X_{i,\tet,\Al}$, $X_{i,\tet,\Sb}$) and hexagonal sites ($X_{i,\hex}$) as well as several split-interstitial configurations oriented along $\left<100\right>$ and $\left<110\right>$. The latter configurations were always observed to relax into a structure in which the bond between the impurity atom and its nearest neighbor is oriented slightly away from the $\left<110\right>$ direction [compare \fig{fig:confs_int_bb}]. In the literature, this distorted structure has been previously referred to as ``bridge'' interstitial, and for simplicity in the following we adopt this convention. We typically calculated formation energies for charge states between $q=-2$ and  $+2$, and where necessary also included charge states $-3$ and $+3$.

We also investigated the formation of defect complexes of group IV and VI elements with aluminum vacancies $V_{\Al}$ and antimony antisites $\Sb_{\Al}$ which are the two most important intrinsic defects.\cite{AbeErhWil08}

\subsection{Computational details}
\label{sect:method_compdetails}

Calculations were performed within density-functional theory (DFT) using the local density approximation (LDA) and the projector augmented wave method \cite{Blo94, KreJou99} as implemented in the Vienna ab-initio simulation pack\-age. \cite{KreHaf93, KreHaf94, KreFur96a, KreFur96b}

In our earlier study \cite{AbeErhWil08} of intrinsic defects in AlSb we employed a variety of supercell sizes in combination with a finite-size scaling procedure in order to mitigate the limitations of the supercell method. Given the large number of different elements, defect structures, and charge states considered in the present work, such a comprehensive approach is, however, beyond our present computational capabilities. We therefore carried out the majority of the calculations for 64-atom supercells, which based on the results of Ref.~\onlinecite{AbeErhWil08} implies that the error in our calculated formation energies due to the supercell approximation is still only on the order of 0.1\,eV. For Brillouin-zone integrations we used a $6\times 6\times 6$ Monkhorst-Pack mesh. \cite{MonPac76} For a small set of configurations, we performed additional calculations using 216-atom cells (see \fig{fig:bands}) and a $4\times 4\times 4$ Monkhorst-Pack $k$-point mesh to verify the viability of our 64-atom cell calculations. We used a plane-wave cutoff energy of 500\,eV, which is imposed by the pseudo-potential for oxygen, but for the sake of consistency was used for all calculations. A Gaussian smearing with a width of $\sigma=0.1\,\eV$ was used to determine the occupation numbers. The ionic positions were optimized until the maximum force was below $20\,\meV/\AAA$. For the charged defect calculations a homogeneous background charge was added to ensure charge neutrality of the entire cell.

For bulk AlSb we obtain a lattice constant of 6.12\,\AA\ in good agreement with the experimental room temperature value of 6.13\,\AA. The band gap is 1.12\,eV, which is lower than the experimental room temperature value of 1.69\,eV as expected for a LDA calculation. A more detailed overview of the bulk properties of Al, Sb, and AlSb that are obtained using the present computational parameters can be found in Ref.~\onlinecite{AbeErhWil08}.

The underestimation of the band gap is a well-known deficiency of DFT within the LDA. In order to overcome this deficiency in Ref.~\onlinecite{AbeErhWil08} we have employed a {\it post-mortem} correction scheme which relies on a separation of valence and conduction band states. Analysis of the electronic structure of the defects in this work reveals that in many cases the defect induces electronic levels in the band gap, which renders the separation into valence, conduction or localized states highly unreliable. Since as a result the application of the correction scheme employed in Ref.~\onlinecite{AbeErhWil08} becomes ambiguous, we have decided to abstain from any band gap corrections in the present work. Instead all results are presented and discussed with respect to the calculated band gap, as done previously by Du. \cite{Du09} For reference we have, however, included the experimental band gap in Figs.~\ref{fig:eform1}, \ref{fig:eform_O}, and \ref{fig:eform_H}.

To compute the binding energies of defect complexes involving aluminum vacancies and antimony antisites, we used the defect formation energies obtained in Ref.~\onlinecite{AbeErhWil08} with for 64-atom cells and without band gap corrections

\subsection{Lattice distortion and effect on charge carrier mobilities}
\label{sect:method_scat}

Defects break the translational symmetry of an ideal crystal and therefore act as scattering centers for electrons and holes traveling through the crystal. The two most important contributions to the scattering cross section of a defect are related to the amount of charge localized at the defect (Coulomb scattering) and the lattice distortions that accompany defect relaxation. Since we are primarily interested in comparing defects that have identical absolute charge states (e.g., $\C_{\Sb}^{-1}$ {\it vs} $\Te_{\Sb}^{+1}$), in the following we focus on the second contribution only. There are many different ways in which one can quantify lattice distortions. We have empirically found that a particularly useful measure is given by
\begin{align}
  M^2 = \left(\int d\vec{r} \left| \nabla_{\vec{r}} (\Delta V) \right| \right)^2,
  \label{eq:relscat}
\end{align}
where $\Delta V$ denotes the potential difference $\Delta V$ between the defective and ideal cells. This quantity includes not only ionic relaxations but also changes in the electronic charge distribution.

In \eq{eq:relscat} we use the potential and not ---for example--- the charge density difference. This choice is motivated by a full fledged computation of the scattering rates \cite{LorErhAbe10} based on Fermi's golden rule for which one needs to evaluate matrix elements of the form $\left<\Psi_f|\Delta V|\Psi_i\right>$.\cite{ChaQue81, EvaZhaJoa05} In that study, \cite{LorErhAbe10} we find that $M^2$ reproduces very well the trends obtained from the scattering rate calculations based on Fermi's golden rule across a wide range of extrinsic defects. Since the computation of $M^2$ is, however, orders of magnitude more efficient, in the following we use the {\it relative scattering strength} $M^2$ to assess the impact of defects on carrier transport. We note that this analysis does not differentiate between electron and hole scattering. It is, however, unusual that a given defect has a significantly different impact on electron and hole scattering. \cite{LorErhAbe10}

\subsection{Self-consistent charge equilibration}
\label{sect:method_chargeneutrality}

In Refs.~\onlinecite{AbeErhWil08} and \onlinecite{ErhAlb08} we have described in detail a self-consistent scheme to determine the concentrations of defects and charge carriers as a function of temperature and chemical environment. Here, we only provide a brief summary of the most important equations. The concentrations of electrons, holes, and charged defects in a material are coupled to each other via the charge neutrality condition
\begin{align}
  n_e + n_A = n_h + n_D,
  \label{eq:charge_neutrality}
\end{align}
where $n_e$ and $n_h$ are the electron and hole concentrations, respectively. They are given by
\begin{align}
  n_e &= \int_{\CBM}^{\infty} D(E) \, f(E,\mu_e) \, dE
  \label{eq:ne}
  \\
  n_h &= \int_{-\infty}^{\VBM} D(E) \, \left[ 1 - f(E,\mu_e) \right] \, dE,
\end{align}
where $D(E)$ denotes the density of states, $f(E,\mu_e)=\left\{1+\exp\left[(E-\mu_e)/k_B T\right]\right\}^{-1}$ is the Fermi-Dirac distribution, $\mu_e$ is the electron chemical potential, $\VBM$ and $\CBM$ indicate the valence band maximum and the conduction band minimum, respectively. The concentrations of acceptors and donors, $n_A$ and $n_D$, can be split into contributions due to intrinsic and extrinsic defects
\begin{align}
  n_A = n_A^{\intt} + n_A^{\ext}
  \quad\text{and}\quad
  n_D = n_D^{\intt} + n_D^{\ext}
\end{align}
The concentration of intrinsic acceptors (donors) is obtained by summing over all intrinsic defects with charge states $q<0$ ($q>0$),
\begin{align}
  n_A^{\intt} = \sum_i^{\text{acceptors}} e \, q_i \, c_{i,q}
  \quad\text{and}\quad
  n_D^{\intt} = \sum_i^{\text{donors}} e \, q_i \, c_{i,q}
  \label{eq:nD}
\end{align}
where $e$ is the unit of charge and the defect concentration is given by \eq{eq:def_conc}.

The concentrations of extrinsic acceptors and donors are given by expressions that are analogous to \eq{eq:nD} but subject to the constraint that the {\em total} concentration of a given impurity or dopant is constant. A condition that can be easily achieved by proper normalization of the defect concentrations. The set of Eqs.~(\ref{eq:charge_neutrality}--\ref{eq:nD}) can be solved self-consistently using an iterative algorithm.


\section{Formation energies and defect structures}
\label{sect:eform}

\subsection{Group IV elements: C, Si, Ge, and Sn}

\begin{figure*}
  \centering
\includegraphics[width=0.9\linewidth]{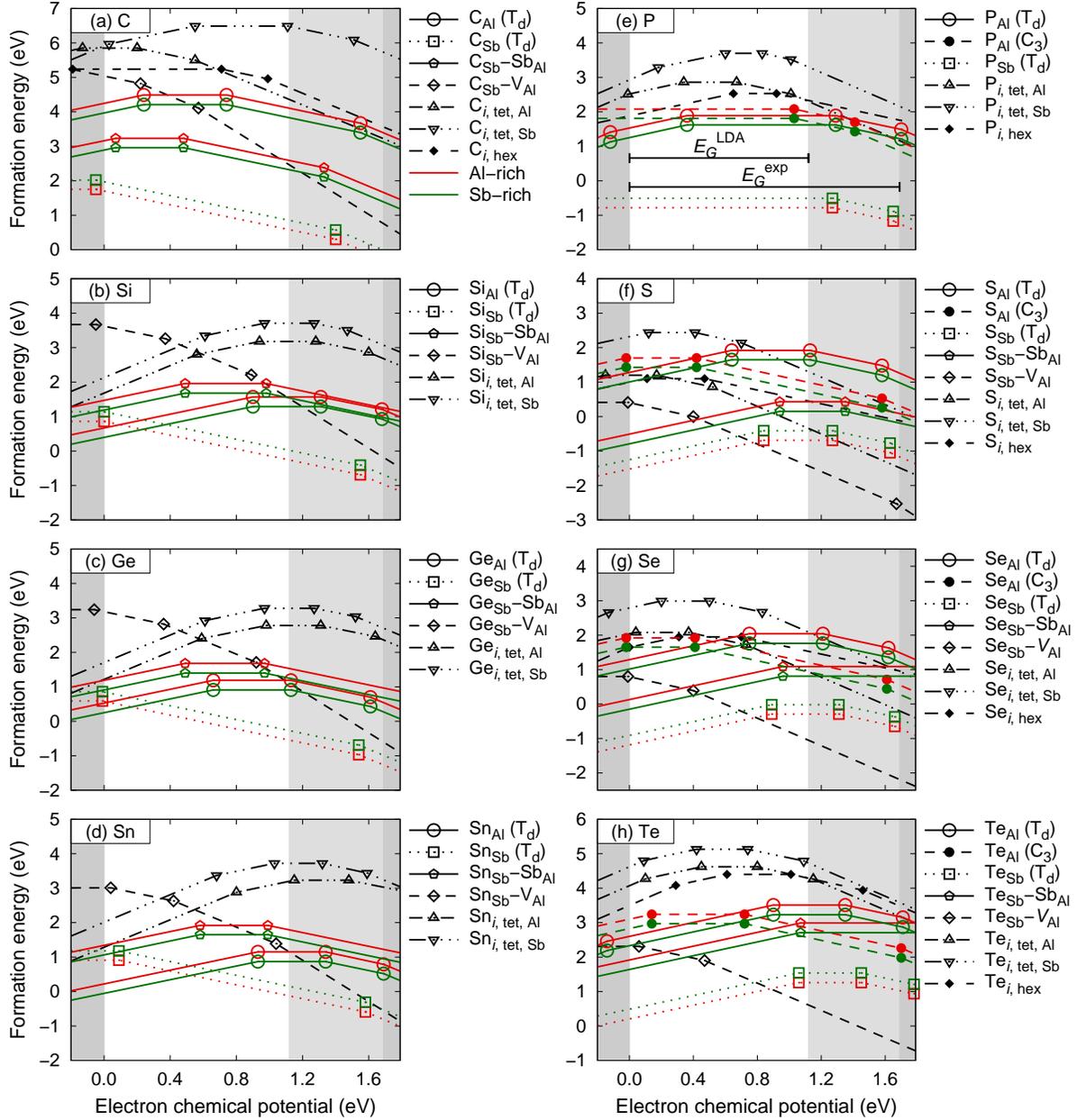}
  \caption{
    (Color online) Formation energies of impurity-related defects for (a-d) group IV (e) P, and (f-h) group VI elements calculated as a function of the electron chemical potential with $\Delta\mu_{\text{imp}}=0\,\eV$ in \eq{eq:eform}. Symbols indicate equilibrium transition levels [see \eq{eq:trans}]. The slope of  the lines corresponds to the charge state. The white region indicates the calculated band gap while the lighter gray region indicates the difference between the experimental and the calculated band gap.
  }
  \label{fig:eform1}
\end{figure*}

\begin{figure*}
  \centering
  \includegraphics[height=1.4in]{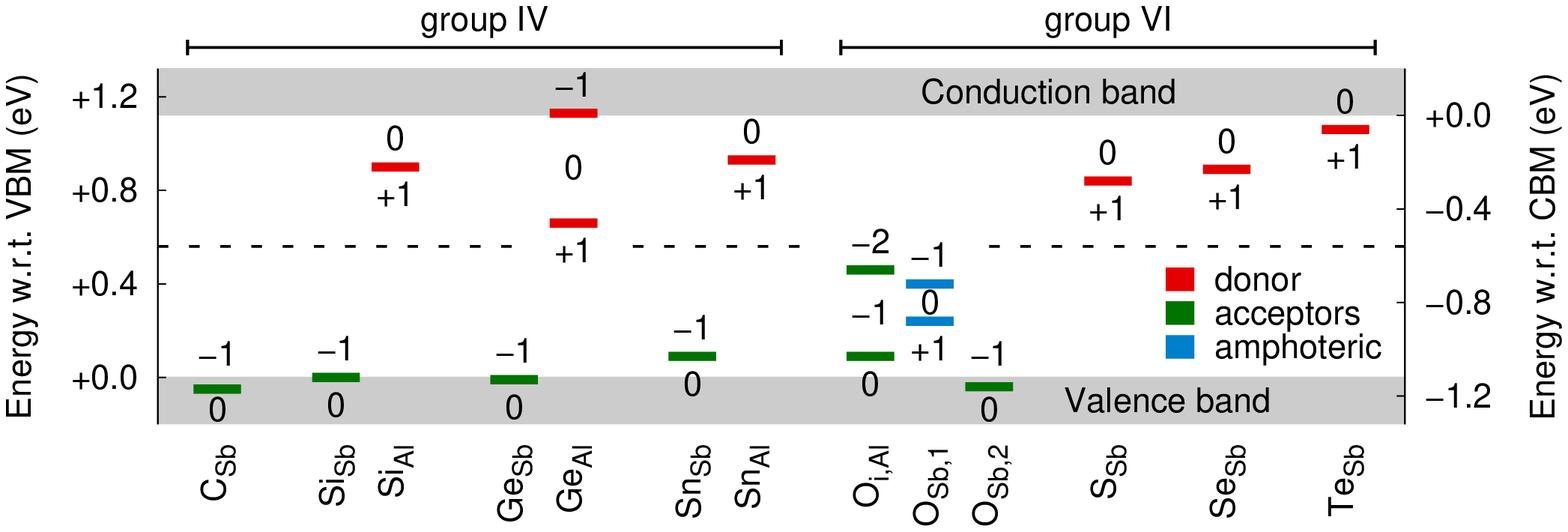}
  \hspace{24pt}
  \includegraphics[height=1.4in]{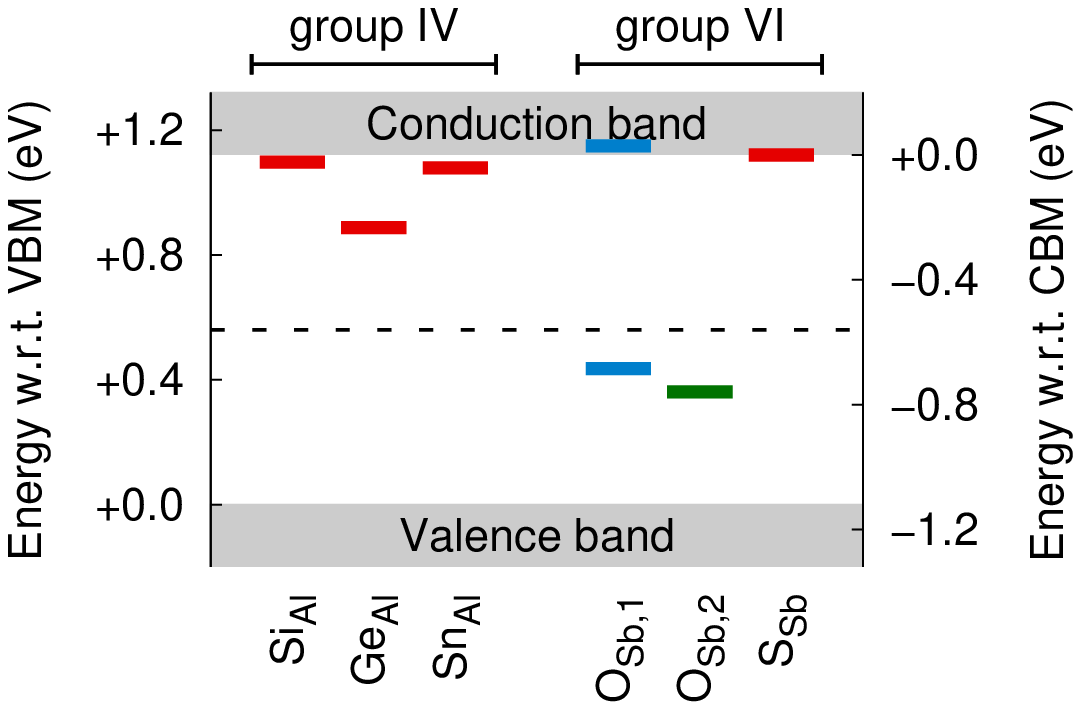}
  \caption{
    (Color online)
    Equilibrium ({\it left}) and electronic ({\it right}) transition levels of the most relevant impurity related defects for group IV and VI elements. The equilibrium transition levels have been extracted from the formation energies using \eq{eq:trans} and correspond to the position of the symbols in Figs.~\ref{fig:eform1} and \ref{fig:eform_O}. The electronic transition levels have been identified using band structure calculations of the type shown in \fig{fig:bands}. $\O_{\Sb,1}$ and $\O_{\Sb,2}$ stand for $\O_{\Sb}(C_{3v})$ and $\O_{\Sb}(\beta-\text{CCBDX}$), respectively. All other substitutional defects have $T_d$ symmetry. The oxygen interstitial has tetrahedral symmetry.
  }
  \label{fig:trans}
\end{figure*}

\begin{figure}
  \centering
\includegraphics[width=0.98\columnwidth]{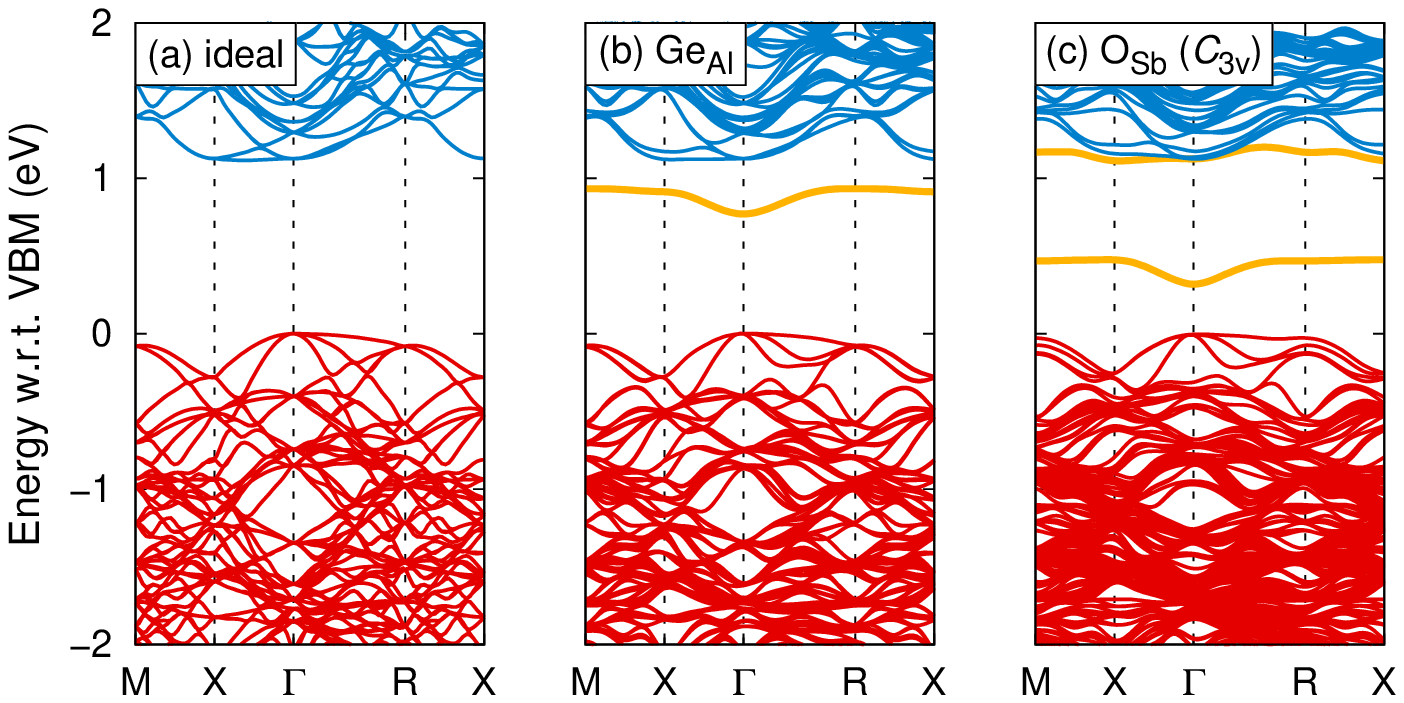}
  \caption{
    (Color online) Band diagrams for (a) an ideal system as well as for systems containing (b) a $\Ge_{\Al}$ defect and (c) a $\O_{\Sb}$ ($\beta$-CCBDX) defect. Conduction and valence bands are indicated by red and blue lines, respectively. Defect states are shown in thick yellow lines. All calculations were carried for 216-atom supercells in charge state $q=0$.
  }
  \label{fig:bands}
\end{figure}

The formation energies of defects involving the first four group IV elements (C, Si, Ge, and Sn) are shown in \fig{fig:eform1}(a--d). In all cases the most important configuration is the substitutional defect on an Sb-site, $X_{\Sb}$ ($X$=C, Si, Ge, or Sn). Regardless of the initial condition, we find the defects to relax into a configuration with $T_d$ symmetry. The $X_{\Sb}$ defects occur predominantly in charge state $q=-1$ and thus act as single acceptors. This observation is in agreement with Hall and resistivity measurements. \cite{Sha62, BenMoo00, McCHalBec01}

Within the error bars of our calculations the $0/-1$ equilibrium transition levels for C, Si, and Ge coincide with the VBM whereas for Sn the equilibrium transition level is located 0.1\,eV above the VBM (compare \fig{fig:trans}). In agreement with our calculations, shallow acceptor levels have been observed for Si by Bennett \etal\ \cite{BenMoo00} while C has been found to act an acceptor by McCluskey \etal. \cite{McCHalBec01} For $X_{\Sb}$ defects, our calculations do not reveal any electronic states inside the band gap that could give rise to strong optical signatures.

For C the substitutional defect on an Al-site $X_{\Al}$, which always adopts a local $T_d$ symmetry, is very high in energy and therefore should not occur in thermodynamic equilibrium. For Si, Ge, and Sn, however, this defect becomes increasingly more important. In fact in the case of Sn, the formation energies for substitutional defects on Al and Sb-sites are very similar. These results are consistent with the experiments by Shaw \cite{Sha62} who observed that Sn can occupy both Al and Sb sites. For Si, Ge, and Sn the calculations predict $X_{\Al}$ to be a donor with a deep $+1/0$ equilibrium transition level (compare \fig{fig:trans}). For Si, Ge, and Sn we also find electronic levels located inside the band gap. This is explicitly demonstrated for the case of $\Ge_{\Al}$ in \fig{fig:bands}(b), which reveals a localized defect state approximately 0.24\,eV below the CBM.

The similar formation energies of $\Sn_{\Al}$ donors and $\Sn_{\Sb}$ acceptors gives rise to an interesting type of amphoteric behavior for Sn-doped AlSb. Usually this property is associated with a single defect type, which undergoes electronic (and usually also structural) changes as a function of charge state. In the case of AlSb:Sn, however, the amphoteric behavior is due to two different defects occupying two different lattice sites that possess opposite electronic characteristics. This leads to an intriguing opportunity for compensation doping that will be explored in \sect{sect:Sn_doping}.

Regarding interstitial defects, both tetrahedral configurations $X_{i,\tet,\Al}$ and $X_{i,\tet,\Sb}$ [Figs.~\ref{fig:confs_int_tet_Al} and \ref{fig:confs_int_tet_Sb}] are found to be local minima on the energy landscape with $X_{i,\tet,\Al}$ being slightly more stable. The hexagonal interstitial $X_{i,\hex}$ is a local minimum for C only, whereas it relaxes to the $X_{i,\tet,\Al}$ configuration for Si, Ge, and Sn. All interstitials are, however, significantly higher in energy than the most stable substitutional defects whence in thermodynamic equilibrium their concentrations should be negligible. Since the migration barriers for substitutional defects are, however, typically very large, it is possible that interstitial configurations play a role with respect to the diffusion of group IV elements.

We also considered defect complexes of group IV impurities with the intrinsic defects $V_{\Al}$ and $\Sb_{\Al}$. The formation energies for $X_{\Sb}-V_{\Al}$ defects which act as acceptors are rather high due to the like charges of the two isolated defects. The $X_{\Sb}-\Sb_{\Al}$ complexes on the other hand display rather small formation energies that are typically less than 1\,eV higher than the formation energy of the most stable configuration. These defects exhibit amphoteric behavior with the donor state dominating over the wider range of the band gap [also see \fig{fig:ebind}(a)].

\subsection{Group V elements: phosphorus}

Phosphorus is included in the present work as a prototypical example for group V elements which are isoelectronic to Sb. The calculated formation energies are shown in \fig{fig:eform1}(e) which shows that P prefers substitution on a Sb-site maintaining $T_d$ symmetry. $\P_{\Sb}$ is neutral throughout the band gap and thus does not introduce any transition levels inside the gap. All other possible configurations are much higher in energy.

\begin{figure*}
  \centering
\includegraphics[width=0.88\linewidth]{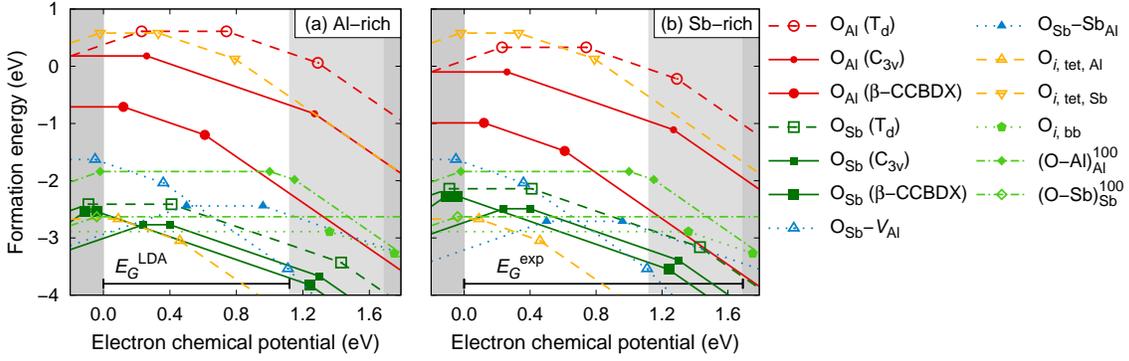}
  \caption{
    (Color online)
    Formation energies of oxygen-related defects under Al-rich and Sb-rich conditions [$\Delta\mu_{\O}=0\,\eV$ in \eq{eq:eform}]. Symbols indicate equilibrium transition levels. The slope of the lines corresponds to the charge state.
  }
  \label{fig:eform_O}
\end{figure*}

\subsection{Group VI elements: S, Se, and Te}
\label{sect:groupVI}

Since oxygen behaves very different from the other group VI elements we first consider the less complex cases of S, Se, and Te. A detailed discussion of oxygen is given in the next section.

For S, Se, and Te the most stable defect configuration for electron chemical potentials over the widest range of the band gap is $X_{\Sb}$ which possesses $T_d$ symmetry and acts as a donor [Figs.~\ref{fig:eform1}(f-h)]. The $+1/0$ equilibrium transition levels for these defect are deep for all three elements (\fig{fig:trans}). While all interstitial configurations considered are local minima, they are always much higher in energy than the respective $X_{\Sb}$ defect. Although there is supporting experimental evidence of the existence of shallow donor levels, \cite{StiBec66, AhlRam68} it is firmly established that the ground-state of these donors for electron chemical potentials near the conduction band minimum is actually a DX-center \cite{ChaCha88, ChaCha89} associated with a deep level.\cite{JosKunKau94, BecWit95, StaWal95} As indicated in \sect{sect:defect_confs}, we have investigated a variety of possible DX-center structures. While we find several of these configurations to be local energy minima for charge state $q=-1$, their formation energies are always higher than the respective $T_d$ configurations in charge state $0$ as long as the electron chemical potential remains within the band gap. This behavior can be attributed to the LDA which in this instance fails to properly capture the energetics of localized {\it vs} delocalized defect states. A detailed discussion is, however, beyond the scope of this article and will be published elsewhere.

Defect complexes of $X_{\Sb}$ with $\Sb_{\Al}$ antisites for the group VI impurities show similar trends as in the case of the group IV impurities being about 1\,eV higher than the isolated $X_{\Sb}$ defects [see \fig{fig:ebind}(c)]. In all cases, these complexes display donor character. In contrast to the group IV elements, however, the $X_{\Sb}-V_{\Al}$ complexes have much lower formation energies resulting from the Coulombic interaction of the oppositely charged isolated defects. In fact, for electron chemical potentials in the upper half of the band gap, this mutual attraction renders the $X_{\Sb}-V_{\Al}$ acceptor complex the most stable defect. The donor character of $X_{\Sb}$ combined with the acceptor character of $X_{\Sb}-V_{\Al}$ could allow for an interesting self-compensation effect. Since this would require that aluminum vacancies remain mobile down to very low temperatures ($\lesssim\,400\,\K$) in order to achieve full equilibrium, such a scenario is, however, unlikely to occur in reality.

\subsection{Group VI elements: oxygen}

Oxygen behaves distinctly different from all the other elements considered so far , in that its interstitial configurations have very low formation energies and there is a multitude of different defect configurations which are all very close in energy (\fig{fig:eform_O}).

Over the widest range of the electron chemical potential, the most stable defect is the tetrahedral interstitial surrounded by Al atoms $\O_{i,\tet,\Al}$ in charge state $q=-2$ [\fig{fig:confs_int_tet_Al}]. This acceptor defect displays transition levels $0.1\,\eV$ ($0/-1$) and $0.5\,\eV$ ($-1/-2$) above the VBM (see \fig{fig:trans}).

For Al-rich conditions and electron chemical potentials in the lower half of the band gap, the lowest energy defect is an interstitial which assumes a bridgelike structure [\fig{fig:confs_int_bb}] and is indicated as $\O_{i,\text{bb}}$ in \fig{fig:eform_O}. It maintains the neutral charge state throughout the band gap. Very close to the VBM the $\O_{\Sb}(C_{3v})$ defect emerges as the ground-state configuration. This defect adopts a configuration with $C_{3v}$ symmetry and is associated with a pronounced lattice distortion [\fig{fig:confs_X_Sb_C3}]. It displays amphoteric characteristics with equilibrium transition levels $0.2\,\eV$ ($+1/0$) and $0.4\,\eV$ ($0/-1$) above the VBM. As shown in Figs.~\ref{fig:trans} and \ref{fig:bands}(c), it also introduces two electronic levels into the band gap that are located 0.45\,eV above the VBM and right at the CBM, respectively.

For Sb-rich conditions and electron chemical potentials in the lower half of the band gap the $\O_{\Sb}-\Sb_{\Al}$ defect complex which exhibits donor character prevail.

Finally, there are several other configurations with formation energies within about $0.3\,\eV$ of the respective ground-state configurations. These include the $(\O-\Sb)_{\Sb}^{\left<100\right>}$ split-interstitial [\fig{fig:confs_XSb_Sb_100}], which is neutral throughout the band gap and electrically inactive, as well as the $\O_{\Sb}(\beta-\text{CCBDX})$ defect [\fig{fig:confs_X_Sb_bCCBDX}].


We observe a large number of oxygen configurations with very similar formation energies, and most of these defects lead to one or more deep equilibrium transition levels. Deep states induce defect-mediated recombination and/or trapping and thereby limit the lifetime of charge carriers. Oxygen incorporation is therefore very detrimental to the operation and energy resolution of a detector device.

\subsection{Hydrogen}

\begin{figure}
  \centering
\includegraphics[width=0.9\linewidth]{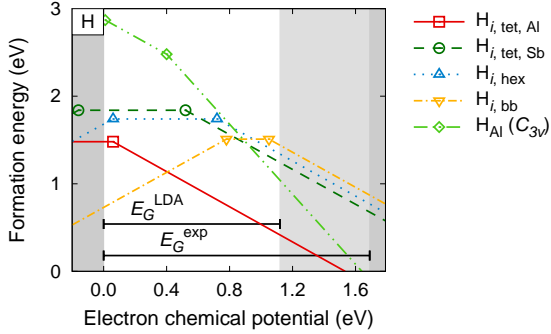}
  \caption{
    (Color online)
    Formation energies of hydrogen-related defects ($\Delta\mu_{\H}=0\,\eV$ in \eq{eq:eform}). Symbols indicate equilibrium transition levels. The slope of the lines corresponds to the charge state [see \eq{eq:eform}].
  }
  \label{fig:eform_H}
\end{figure}

For hydrogen, we focused on interstitial configurations and considered both tetrahedral sites and the bridge bond configuration [\fig{fig:confs_int_bb_H}]. The results are shown in \fig{fig:eform_H}. For electron chemical potentials in the lower half of the band gap, the bridge bond configuration is the energetically most favorable one. It acts as a deep donor with an equilibrium transition level $0.3\,\eV$ below the CBM. For electron chemical potentials in the upper half of the band gap, the $\H_{i,\tet,\Al}$ defect prevails which behaves as an acceptor with an equilibrium transition level $0.1\,\eV$ above the VBM. We also note that although the formation energies for $\H_{i,\text{bb}}$ and $\H_{i,\tet,\Al}$ in the neutral charge state are virtually identical, the structures are nonetheless distinct.

\subsection{Defect complexes}
\label{sect:complexes}

\begin{figure*}
  \centering
\includegraphics[width=0.42\linewidth]{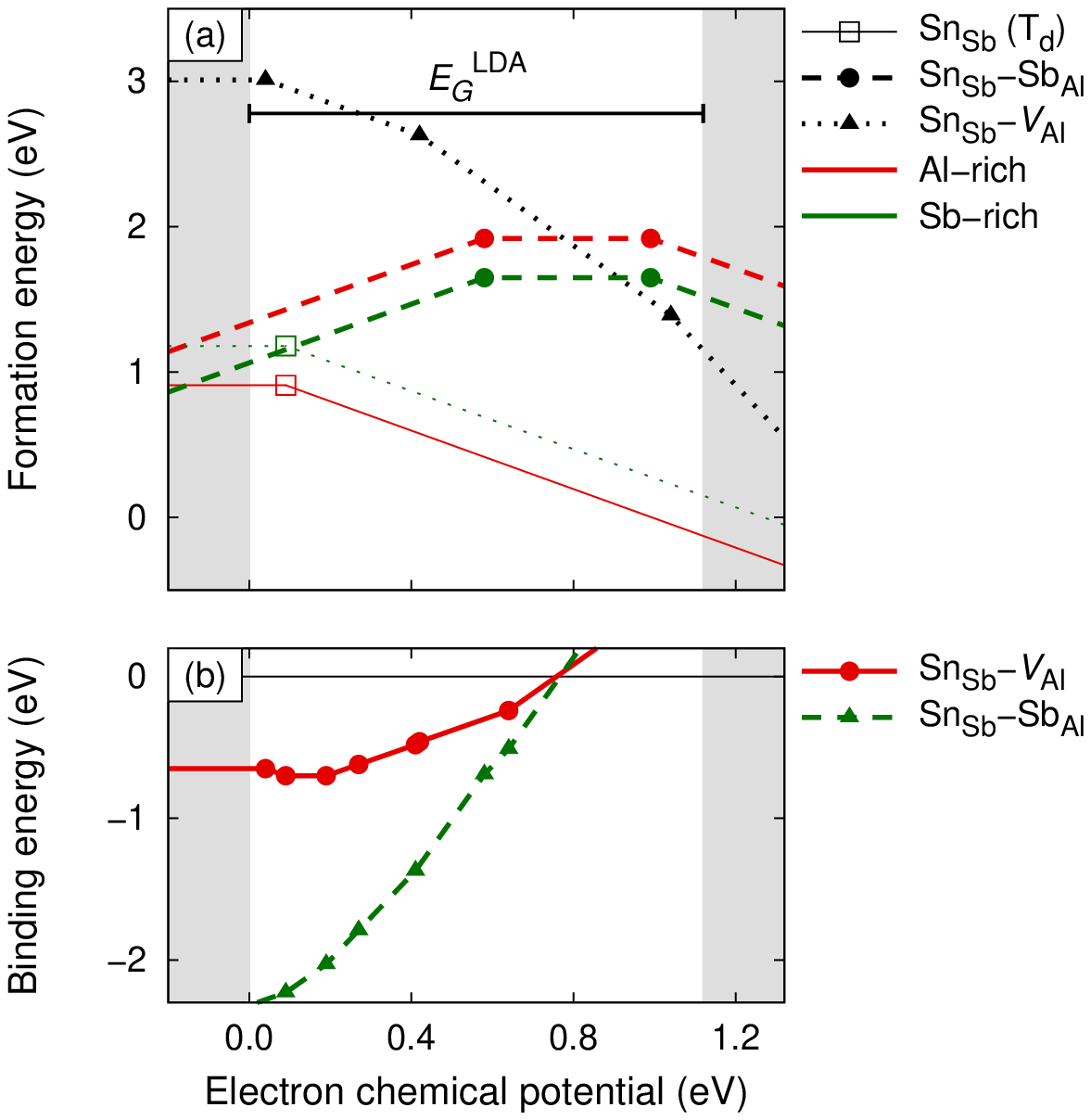}
  \hspace{0.2in}
\includegraphics[width=0.42\linewidth]{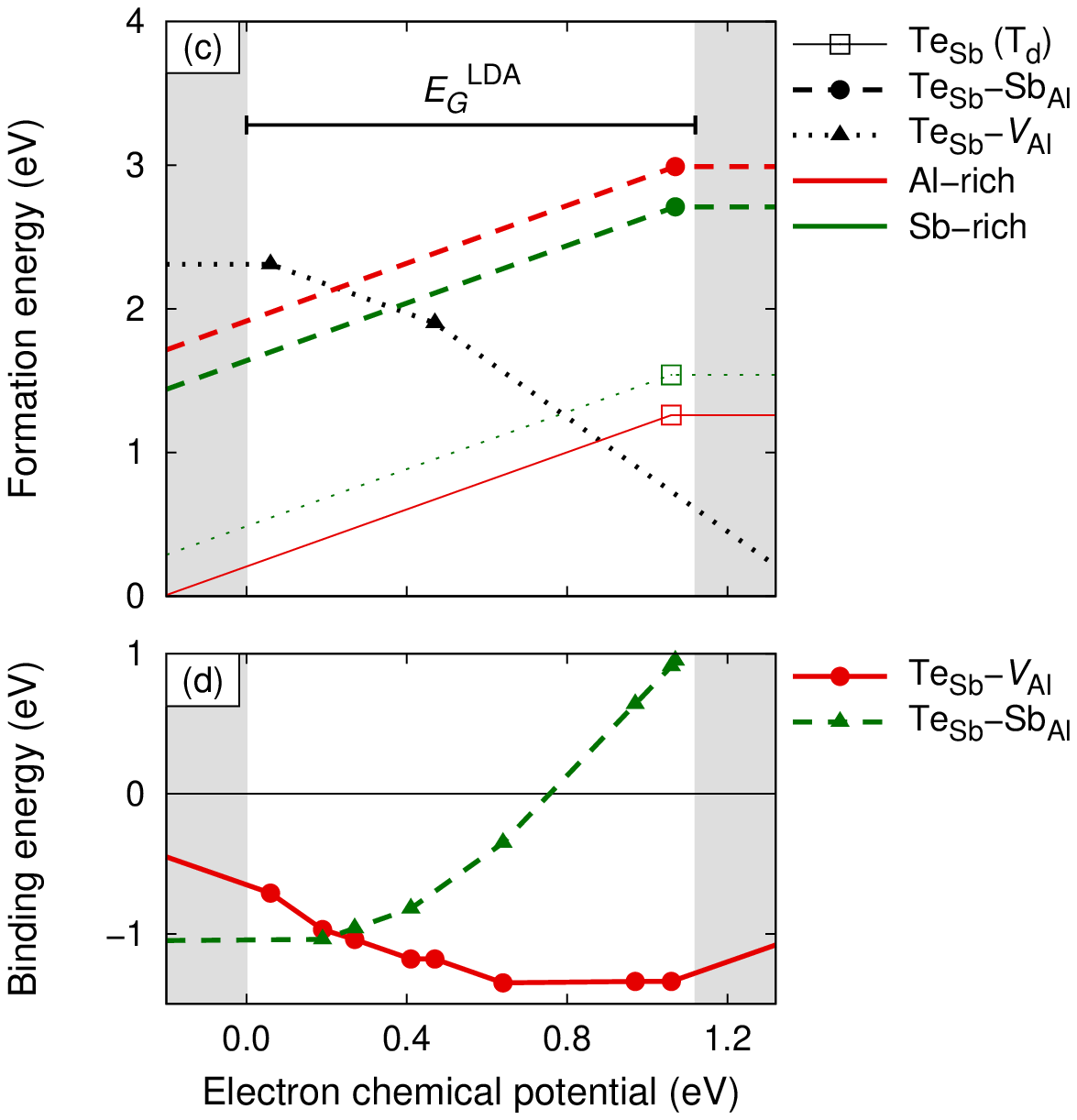}
  \caption{
    (Color online) (a,c) Formation  and (b,d) binding energies of defect complexes of intrinsic defects with (a,b) Sn and (c,d) Te. The two cases are exemplifying the situation for group IV and VI elements, respectively.
  }
  \label{fig:ebind}
\end{figure*}

\begin{figure}
  \centering
\includegraphics[width=0.75\linewidth]{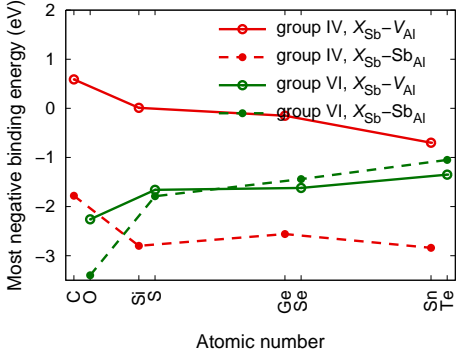}
  \caption{
    (Color online) Most negative binding energies for $X_{\Sb}-V_{\Al}$ and $X_{\Sb}-\Sb_{\Al}$ complexes where $X$ is a group IV or VI element. Following \eq{eq:ebind} negative binding energies indicate a driving force for defect complex formation.
  }
  \label{fig:ebind_max}
\end{figure}

In this section, we consider the interaction of the two most important intrinsic defects \cite{AbeErhWil08} $\Sb_{\Al}$ and $V_{\Al}$ with group IV and VI impurities. In the preceding sections, we have already seen that these defect complexes can have very low formation energies, in particular in the case of group VI elements, where for certain electron chemical potentials they are the most stable defect configuration.

In \fig{fig:ebind}, we show the binding energies and the respective formation energies of Sn and Te-complexes that are prototypical for group IV and VI elements, respectively. To illustrate the trends we show in \fig{fig:ebind_max} the most negative binding energies (indicating the strongest driving force for defect complex formation) for all group IV and VI impurities considered.

In the case of Sn, both types of complexes display strong binding for electron chemical potentials in the lower two-thirds of the band gap with binding energies as negative as $-2.2\,\eV$ for $\Sn_{\Sb}-\Sb_{\Al}$ at the valence band maximum. On the other hand, for electron chemical potentials in the upper third of the band gap, the interaction is strongly repulsive. These trends hold for the other group IV elements as well with the exception of the $\C_{\Sb}-V_{\Al}$ complex which is never bound. In general, the vacancy complexes have high formation energies and act as acceptors, while the antisite complexes have rather low formation energies near the valence band edge and display mostly donor characteristics.

As exemplified by Te, the trends described above for $X_{\Sb}-\Sb_{\Al}$ prevails also for group VI elements. With respect to $X_{\Sb}-V_{\Al}$ complexes with group VI impurities, we obtain binding energies which are consistently negative throughout the band gap. As discussed in \sect{sect:groupVI}, the vacancy complexes act as acceptors, have relatively low formation energies, and are actually the most stable defect for electron chemical potentials near the CBM.

\section{Relative scattering strengths}
\label{sect:relscat}

\begin{figure*}
  \setlength{\myhgt}{2.1in}
  \centering
\includegraphics[height=\myhgt]{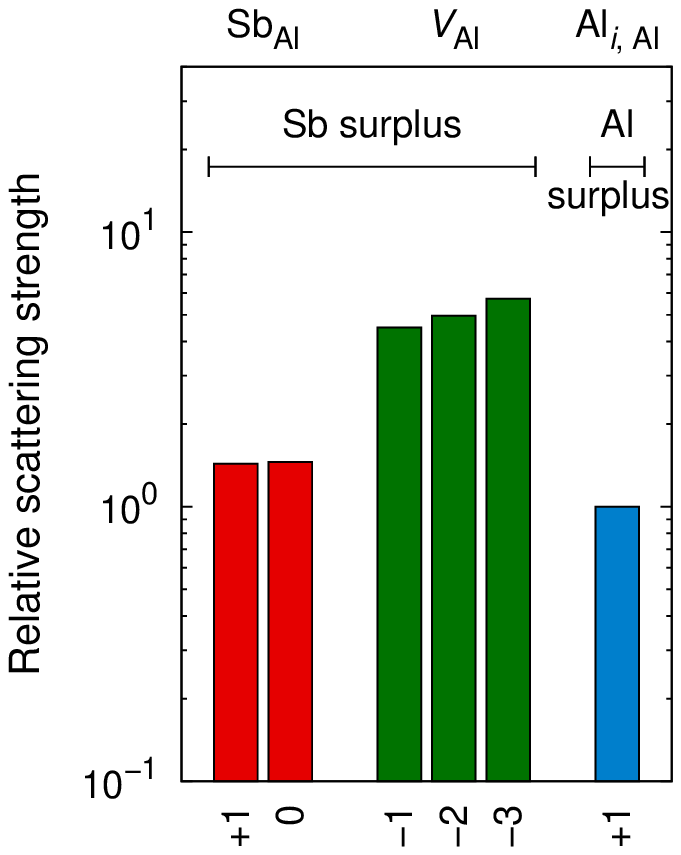}
\includegraphics[height=\myhgt]{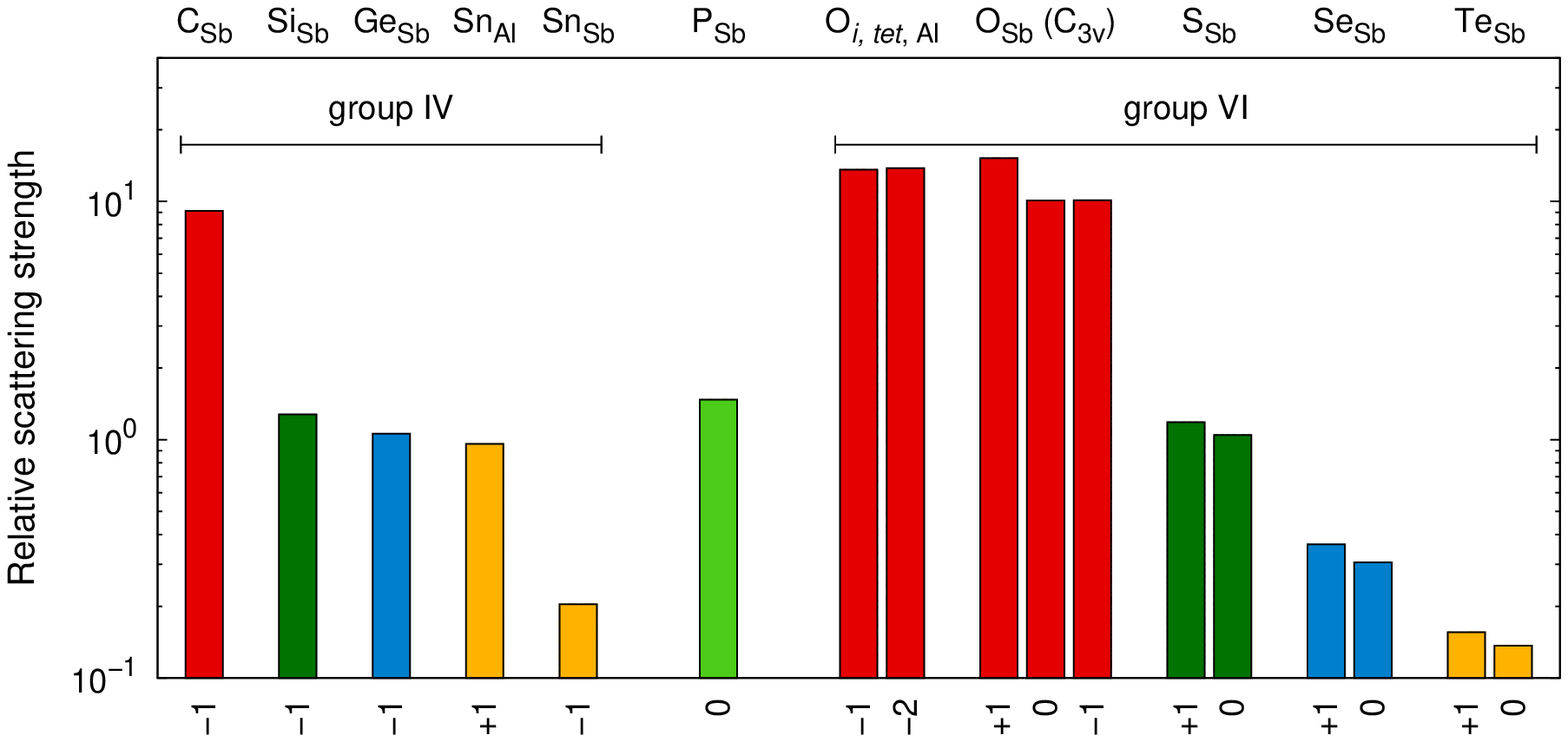}
  \caption{
    (Color online) Relative carrier scattering strengths for the most prevalent intrinsic (left) and extrinsic (right) defects calculated using \eq{eq:relscat}. The substitutional defects all have $T_d$ symmetry with the exception of oxygen for which the results for the $C_{3v}$ configuration are shown. In the left panel, red arrows mark the most important charge states for the intrinsic defects (Ref.~\onlinecite{AbeErhWil08}).
  }
  \label{fig:relscat}
\end{figure*}

The relative carrier scattering strengths calculated using \eq{eq:relscat} for the most important intrinsic and extrinsic defects vary over several orders of magnitude as shown in \fig{fig:relscat}. While the absolute values of the relative scattering strength bear no meaning, they still allow us to compare different extrinsic and intrinsic defects with respect to their impact on charge carrier mobility.\cite{LorErhAbe10}

Figure~\ref{fig:relscat} clearly shows that oxygen induces by far the strongest carrier scattering and thus has the most detrimental impact on charge carrier mobilities, with carbon also showing a very high scattering strength. The relative scattering strengths for these impurities are about one order of magnitude larger than for the aluminum vacancy, which is the most detrimental intrinsic defect.

For the elements within the same group of the periodic table, the scattering strength generally decreases with increasing atomic number. This effect is owed to the diminishing size mismatch between the impurity atom and Sb, for which most impurities substitute, leading to decreased lattice distortion upon substitution. As a result, the elements that have the least negative effect on carrier mobility, are Sn, Se, and particularly Te. In fact, many of the substitutional impurities scatter carriers less strongly than the intrinsic defects. These elements are therefore excellent candidates for doping the material in order to achieve carrier compensation with minimal impact on mobility, an opportunity that is explored in the following section.

\section{Charge carrier concentrations}
\label{sect:dneh}

In the previous sections we have determined the formation energies of a wide range of extrinsic defects and discussed which of them act as donors or acceptors. Furthermore, we have obtained their relative scattering rates, which enabled us to identify the defects which are, respectively, the least and most detrimental to the electron and hole mobilities in the material. Now, we are in a position to combine these data to predict net charge carrier concentrations and to determine conditions for obtaining optimal material. For radiation detection, the objective is to obtain a material which features maximal charge carrier mobilities, which requires small scattering rates, and very high resistivity, which requires low net charge carrier concentrations.

In the analysis that follows, we consider four important cases. First, we analyze the pure material (containing only intrinsic defects). Then, we include in our model unintentional impurities from the growth process, followed by analysis of carrier compensation with intentional Te doping. Finally, we consider the case of Sn doping, which exhibits unique behavior due to its amphoteric character.

\subsection{Pure material}

\begin{figure}
  \centering
\includegraphics[width=0.9\columnwidth]{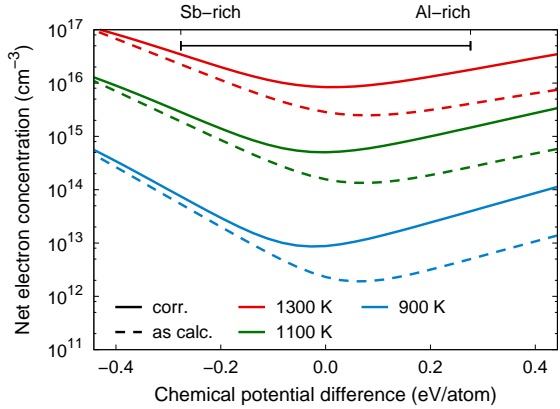}
  \caption{
    Net electron concentration for pure AlSb at three different temperatures using band gap corrected (solid lines) and as-calculated (dashed lines) formation energies assuming full equilibrium. The horizontal bar indicates the range of chemical potentials that are accessible based on the calculated compound formation energy.
  }
  \label{fig:dneh_ideal}
\end{figure}

In Ref.~\onlinecite{AbeErhWil08}, we employed the self-consistent charge equilibration scheme described in \sect{sect:method_chargeneutrality} to determine the concentrations of intrinsic defects and charge carriers in pure AlSb (no impurities) using defect formation energies calculated from DFT, which had been corrected for the band gap error. As discussed in \sect{sect:method_compdetails}, in the present work we did not employ any such corrections. Before we can apply the self-consistent charge equilibration scheme to study the behavior of extrinsically doped AlSb, we therefore have to discuss to which extent the band gap error influences the results. According to \eq{eq:ne} the intrinsic electron and hole concentrations depend exponentially on the band gap. To be consistent with the calculations discussed above, we now have to use the LDA band gap and the corresponding formation energies for the intrinsic defects. The {\em absolute} carrier concentrations are therefore no longer comparable to our calculations in Ref.~\onlinecite{AbeErhWil08}. For the present purpose, we are, however, interested not in absolute carrier concentrations but in {\em net} charge carrier concentrations, i.e. the difference between the concentrations of electron and holes,
\begin{align}
  \Delta n &= n_e - n_h.
  \label{eq:dneh}
\end{align}
The impact of the band gap error on $\Delta n$ is not immediately obvious. In \fig{fig:dneh_ideal}, we therefore compare the net electron concentrations for pure AlSb using as-calculated formation energies as well as formation energies, which have been corrected for the band gap error. We find that while the actual values deviate somewhat, the conduction type ($n$ {\it vs} $p$), chemical potential dependence, and the order of magnitude of $\Delta n$ are very similar. This finding enables us to use the as-calculated formation energies to compute net charge carrier concentrations for AlSb with extrinsic defects. Figure~\ref{fig:dneh_ideal} also shows that pure AlSb exhibits $n$-type behavior over the entire range of the chemical potential difference. In contrast, as-grown AlSb has been found to exhibit $p$-type conductivity. In the following section, we show that this apparent discrepancy is caused by the accidental incorporation of impurities during growth.

\subsection{Unintentional background doping}

In the experimental growth process, certain impurities are impossible to avoid. In particular this includes oxygen and carbon.\cite{McCHalBec01, ErhAbeStu09} Oxygen can be incorporated from the gas phase or the components of the growth chamber and is also present in the raw starting materials. Carbon enters because graphite susceptors are used to heat the melt.\footnote{Additional impurities have been measured in high-purity samples using secondary ion mass spectroscopy including S, Si, and P. These impurities can, however, be fairly well controlled experimentally and according to our calculations in \sect{sect:relscat} are also not as detrimental for carrier mobility as C and O.} Using a highly optimized growth process, it is possible to reduce the concentration of oxygen to about $10^{15}\,\cm^{-3}$ and the concentration of carbon to about $10^{17}\,\cm^{-3}$.\cite{ErhAbeStu09}


In \fig{fig:dneh_CO}, we show the net charge carrier concentrations for AlSb with $c(\C) = 10^{17}\,\cm^{-3}$ and $c(\O)=10^{15}\,\cm^{-3}$. In fact, these concentrations cause the conductivity to switch from $n$-type to $p$-type, which resolves the apparent discrepancy between theory and experiment mentioned above. The inversion of the conductivity-type is indeed expected since in \sect{sect:eform} both of these impurities have been identified as acceptors.

\subsection{Compensation with Te}

\begin{figure}
  \centering
\includegraphics[width=0.9\columnwidth]{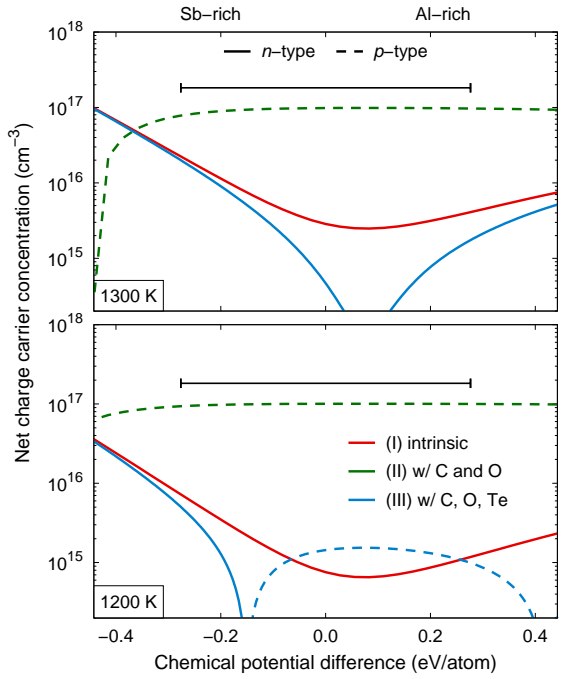}
  \caption{
    Net charge carrier concentration for (I) pure AlSb, (II) AlSb with $c(\C) = 10^{17}\,\cm^{-3}$ and $c(\O)=10^{15}\,\cm^{-3}$, and (III) AlSb containing $c(\C) = 10^{17}\,\cm^{-3}$, $c(\O)=10^{15}\,\cm^{-3}$, and $c(\Te)=10^{17}\,\cm^{-3}$ for two different temperatures assuming full equilibration and using as-calculated formation energies. If $\Delta n$ in \eq{eq:dneh} is positive (negative) as shown by the solid (dashed) lines the material exhibits $n$-type ($p$-type) conductivity. The horizontal bar indicates the range of chemical potentials that are accessible based on the calculated compound formation energy.
  }
  \label{fig:dneh_CO}
  \label{fig:dneh_COTe}
\end{figure}

Since the background doping by C and O renders the material $p$-type, compensation requires addition of donor ions. To simultaneously achieve high carrier mobilities and low net carrier concentrations, the dopant should possess a scattering potential which is as small as possible. According to Figs.~\ref{fig:eform1} and \ref{fig:relscat} Te is an ideal candidate based on these criteria. In \fig{fig:dneh_COTe} we show the effect of adding Te at a concentration of $10^{17}\,\cm^{-3}$ to AlSb containing C and O impurities. We find that by using Te doping in this concentration range the net charge carrier concentration can be tuned very efficiently. The net carrier concentration changes as a function of the chemical potential difference between Al and Sb, with a minimum close to the Sb-rich limit (typical annealing condition). Intentional doping with Te therefore enables us to compensate the background doping due to C and O impurities while minimizing the detrimental impact on the carrier mobilities. In fact, using these data as a guidance for the experimental growth process led to a dramatic improvement of the material for radiation detection, which is reported elsewhere. \cite{ErhAbeStu09}

\begin{figure}
  \centering
\includegraphics[width=0.9\columnwidth]{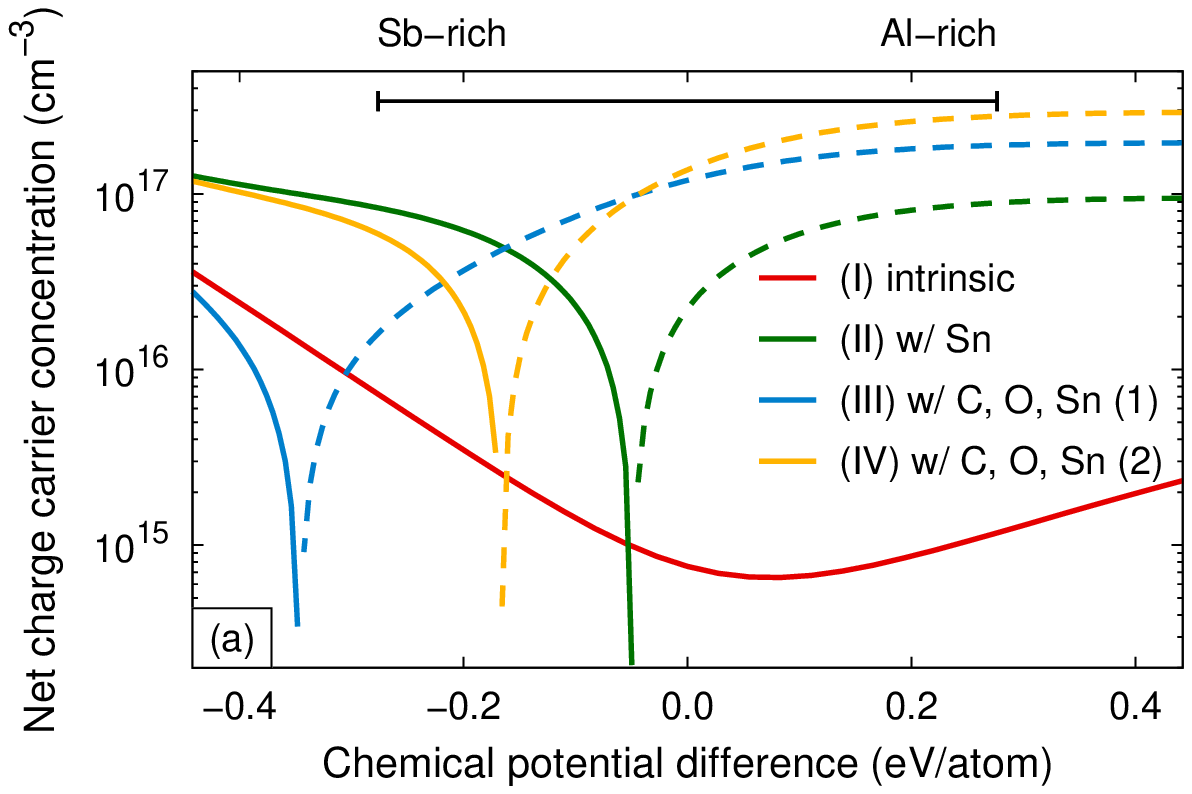}
\includegraphics[width=0.9\columnwidth]{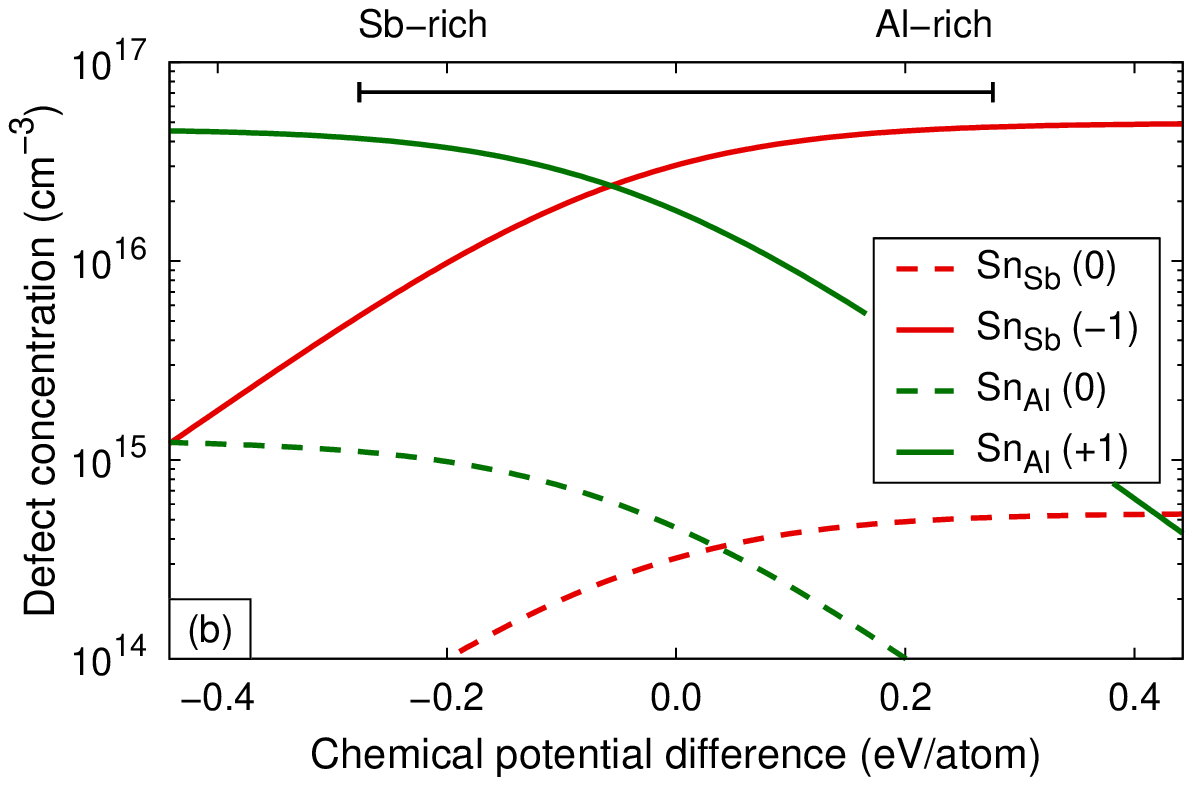}
  \caption{
    (a) Net charge carrier concentration for (I) pure AlSb, (II) AlSb with $c(\Sn) = 10^{17}\,\cm^{-3}$, (III) AlSb with $c(\C) = 10^{17}\,\cm^{-3}$, $c(\O)=10^{15}\,\cm^{-3}$, and $c(\Sn)=10^{17}\,\cm^{-3}$, and (IV) AlSb with $c(\C) = 10^{17}\,\cm^{-3}$, $c(\O)=10^{15}\,\cm^{-3}$, and $c(\Sn)=2\times 10^{17}\,\cm^{-3}$ at a temperature of 1200\,K assuming full equilibration and using as-calculated formation energies. Note that in cases (II--IV) going from Sb-rich to Al-rich conditions changes the conductivity type from $n$ to $p$. (b) Defect concentrations for case (II) in (a) revealing that the inversion of the conductivity type observed in Sn-doped AlSb is caused by a shift from $\Sn_{\Al}^{+1}$ to $\Sn_{\Sb}^{-1}$. The horizontal bars indicates the range of chemical potentials which are accessible based on the calculated compound formation energy.
  }
  \label{fig:dneh_COSn}
\end{figure}

\subsection{Ambipolar doping with Sn}
\label{sect:Sn_doping}

According to \fig{fig:eform1}, Sn can be incorporated both on the Sb and the Al sublattice, which gives rise to acceptor and donor-type defects, respectively. Also, \fig{fig:relscat} shows that both of these defects are weak scatterers compared to the aluminum vacancy. Figure~\ref{fig:dneh_COSn}(a) illustrates the effect of Sn on the net charge carrier concentration. In thermodynamic equilibrium and for a Sn concentration of about $10^{17}\,\cm^{-3}$, the material is $n$-type if grown under Sb-rich conditions and $p$-type if grown under Al-rich conditions [case II in \fig{fig:dneh_COSn}(a)]. In the presence of C and O impurities this behavior is retained although the transition point is shifted toward Sb-rich conditions [case C in \fig{fig:dneh_COSn}(a)]. By increasing the Sn concentration the minimum in $\Delta n$ can be shifted back toward the impurity-free case [case IV in \fig{fig:dneh_COSn}(a)]. Figure~\ref{fig:dneh_COSn}(b) reveals that the origin of this behavior is the subtle equilibrium between $\Sn_{\Sb}^{-1}$ and $\Sn_{\Al}^{+1}$ defects, which act as acceptors and donors, respectively.

We suggest that this intriguing property of Sn-doped AlSb can be exploited to achieve self-compensation. In a typical experiment, an AlSb single crystal is grown under Al-rich conditions close to the melting point of 1330\,K. The growth process is sufficiently slow to assume that we are near thermodynamic equilibrium, which implies that the vast majority of Sn occupies Sb-sites. The growth step is usually followed by a high-temperature anneal at about 1200\,K under Sb-rich conditions. This inversion of chemical conditions leads to a strong driving force for redistributing Sn from the Sb to the Al sublattice. This conversion can be mediated by aluminum vacancies which at this temperature are both very mobile \cite{VAl_mobility} and exist in high concentrations.\cite{AbeErhWil08} By carefully controlling the annealing time it should thus be possible to achieve a $\Sn_{\Sb}/\Sn_{\Al}$ ratio close to one. At this point the material would be perfectly compensated due to the opposing effects of $\Sn_{\Sb}$ and $\Sn_{\Al}$ defects. Due to the moderate scattering strength of Sn-related defects such a material would exhibit high carrier mobilities in conjunction with high resistivity, as needed for the radiation detection applications.

\section{Conclusions}

In this work we have investigated the thermodynamic and electronic properties of group IV and group VI impurities as well as P and H in AlSb using density-functional theory calculations. First, we computed the formation energies for a wide range of possible defect configurations and charge states to identify the most important defect configurations, their electronic characteristics, and their equilibrium and electronic transition levels. We found that C, Si, and Ge preferentially substitute for Sb, act as acceptors and exhibit shallow transition levels. Sn, however, can substitute for Sb and Al and displays amphoteric behavior.

With regard to group VI elements, S, Se, and Te substitute for Sb and have deep donor levels. Oxygen in contrast can be incorporated in several different configurations that display a variety of electronic characteristics and transition levels, predominantly of the acceptor type. For the widest range of electron chemical potentials a tetrahedral interstitial configuration is the lowest in energy.

Phosphorus substitutes for Sb and is electrically inactive while H is incorporated on interstitial sites and can act as acceptor or donor depending on conditions.

With respect to defect complexes, our calculations show that group IV elements bind to both aluminum vacancies ($V_{\Al}$) and Sb antisites ($\Sb_{\Al}$) for electron chemical potentials in the lower two-thirds of the band gap. (The $\C_{\Sb}-V_{\Al}$ complex is an exception and is never bound). The vacancy complexes have high formation energies and act as acceptors while the antisite complexes have rather low formation energies near the valence band edge and display mostly donor characteristics. Similar trends hold for group VI elements but in this case the vacancy complexes show negative binding energies over the entire range of the band gap. In addition, the formation energies of antisite complexes are very low near the conduction band edge, which for S, Se, and Te renders them even more stable than the individual $X_{\Sb}$ defects in a small range of electron chemical potentials in the upper third of the band gap.

To characterize the effect of different impurities on the charge carrier mobilities, we calculated relative scattering strengths. This analysis revealed that C and O have by far the most detrimental impact on carrier transport. The scattering strengths for Sn, Se, and Te, however, are similar or even smaller than for the most important intrinsic defects. Tellurium, in particular, is therefore an excellent candidate for compensation doping the material.

The interplay of electronic characteristics and impurity concentrations was investigated using a self-consistent charge equilibration scheme that allowed us to compute net charge carrier concentrations. While pure material was found to exhibit $n$-type conductivity, the incorporation of C and O leads to $p$-type material. This observation is consistent with experimental measurements which show $p$-type conductivity in as-grown samples which contain an unintentional background concentration of C and O impurities. Our calculations furthermore revealed how Te can be used to compensate these impurities in order to achieve minimal net charge carrier concentrations. Finally, we discussed how the amphoteric character of Sn could be used to achieve self-compensated material via controlled annealing.

\begin{acknowledgments}
This work performed under the auspices of the U.S. Department of Energy by Lawrence Livermore National Laboratory under Contract No. DE-AC52-07NA27344 with support from the Laboratory Directed Research and Development Program and from the National Nuclear Security Administration Office of Nonproliferation Research and Development (NA-22).
\end{acknowledgments}

\end{document}